\pdfoutput=1

\documentclass[preprint]{aastex631}
\usepackage[utf8]{inputenc}

\newcommand{\bvec}[1]{ \mbox{\boldmath$#1$} }

\usepackage{graphicx}
\usepackage{amssymb}
\usepackage{textgreek}
\usepackage{float}
\usepackage{textcomp,gensymb}
\DeclareUnicodeCharacter{0093}{\^A}

\accepted{May 24, 2023}

\submitjournal{ApJ}

\shorttitle{Atmospheric Gravity Waves}
\shortauthors{Vesa et al.}
\begin{document}

\title{Multi-Height Observations of Atmospheric Gravity Waves at Solar Disk Center}

\correspondingauthor{Oana Vesa}
\email{ovesa@nmsu.edu}

\author[0000-0001-6754-1520]{Oana Vesa}
\affiliation{Department of Astronomy, \\
New Mexico State University, \\
P.O. Box 30001, MSC 4500, \\
Las Cruces, NM 88003-8001, USA}

\author[0000-0001-9659-7486]{Jason Jackiewicz}
\affiliation{Department of Astronomy, \\
New Mexico State University, \\
P.O. Box 30001, MSC 4500, \\
Las Cruces, NM 88003-8001, USA}

\author[0000-0001-8016-0001]{Kevin Reardon}
\affiliation{National Solar Observatory, \\
Boulder, CO 80303, USA}





\begin{abstract}
Atmospheric gravity waves (AGWs) are low-frequency, buoyancy-driven waves that are generated by turbulent convection and propagate obliquely throughout the solar atmosphere. Their proposed energy contribution to the lower solar atmosphere and sensitivity to atmospheric parameters (e.g. magnetic fields and radiative damping) highlight their diagnostic potential. We investigate AGWs near a quiet Sun disk center region using multi-wavelength data from the Interferometric BIdimensional Spectrometer (IBIS) and the Solar Dynamics Observatory (SDO). These observations showcase the complex wave behavior present in the entire acoustic-gravity wave spectrum. Using Fourier spectral analysis and local helioseismology techniques on simultaneously observed line core Doppler velocity and intensity fluctuations, we study both the vertical and horizontal properties of AGWs.Propagating AGWs with perpendicular group and phase velocities are detected at the expected temporal and spatial scales throughout the lower solar atmosphere. We also find previously unobserved, varied phase difference distributions among our velocity and intensity diagnostic combinations. Time-distance analysis indicates that AGWs travel with an average group speed of 4.5\,km\,s$^{-1}$, which is only partially described by a simple simulation suggesting that high-frequency AGWs dominate the signal. Analysis of the  median magnetic field (4.2\,G) suggests that propagating AGWs are not significantly affected by quiet Sun photospheric magnetic fields. Our results illustrate the importance of multi-height observations and the necessity of future work to properly characterize this observed behavior.
\end{abstract}



\section{Introduction} \label{sec:introduction}
The solar atmosphere is a conducive environment for the generation and propagation of a plethora of wave motions that co-exist and interact with one another, including atmospheric gravity waves (AGWs). To avoid confusion with the interior standing g-modes (or internal gravity waves), we use the term AGWs to denote propagating gravity waves throughout the solar atmosphere. Along with other commonly studied waves, AGWs might play an important role in transporting energy and heating the lower solar atmosphere. AGWs are buoyancy-driven waves with gravity acting as their restoring force. These low-frequency waves are believed to be excited stochastically by turbulent convection below the stably stratified surface. The low temporal frequencies (1-4\,mHz), short vertical wavelengths, and transverse propagation has made it difficult to observationally measure and track these waves. A detailed understanding of these waves would not only provide further insight into their behavior but also a window into the complex dynamics of the solar atmosphere and their interactions with the magnetic field.

Studies of AGWs in the field of solar physics began with \citet{1963_Whitaker} and their proposed contribution to the coronal heating problem. Since then, phase difference analysis and other diagnostics using intensity and velocity fluctuations sampled simultaneously at different atmospheric heights near disk center on the quiet Sun \citep[for instance,][]{1984_Staiger_Mattig_Schieder_Deubner,1989_Deubner, 1991_Komm_Mattig_Nesis,1991_Bonnet_Marquez_Vazquez_Woehl, 1992_Deubner,1993_Kneer_vonUexkuel,1997_Straus_Bonaccini, 2001_Krijger_Rutten,2003_Rutten,2008_stodilka,2008_Straus, 2011_kneer_Bello,2014_Nagashima,2021_Calchetti_Jefferies_Fleck_Berrilli_etal} have provided observational evidence for the existence of AGWs from the low photosphere to the low chromosphere.

An extensive theoretical framework on the characterization and energy dissipation of these waves with radiative damping was established in the seminal works of \citet{1981_Mihalas_Toomre,1982_Mihalas_Toomre}. These waves are expected to be affected by severe radiative damping in the lower photosphere which suppresses their propagation \citep[see][]{1966_Souffrin, 1967_Schatzman_Souffrin, 1970_Stix}. However, the observational detection of these waves throughout the lower solar atmosphere indicates that they have enough energy to overcome this dissipative process. With previous observational studies of AGWs focused near disk center, we lack insight into their transverse nature. Numerical simulations by \cite{1981_Mihalas_Toomre} show that AGWs can reach horizontal velocities of 5$-$6\,km\,s$^{-1}$ in comparison to vertical velocities of 1$-$2\,km\,s$^{-1}$, indicating the importance of their horizontal propagation.

In recent years, the effects of the magnetic field on these waves have been explored \citep[for example,][]{2009_Newington_Cally,2011_Newington_Cally,2016_Hague_Erdelyi,2017_Vigeesh,2019_Vigeesh,2020_Vigeesh_Roth,2021_Vigeesh}, indicating their potential as diagnostics for the average magnetic field. Realistic numerical simulations of the magnetized solar atmosphere carried out using CO\textsuperscript{5}BOLD code \citep{2012_Freytag_COBOLD} by \citet{2017_Vigeesh,2019_Vigeesh} and \citet{2021_Vigeesh} demonstrate that AGWs are generated abundantly and propagate irrespective of the field strength and strong radiative damping in the low photosphere. These simulations show that the properties of the magnetic field in the upper photosphere can significantly modify their propagation. Regions of strong, vertical magnetic fields could act to suppress the upward propagation of AGWs \citep{2017_Vigeesh} while horizontal magnetic fields would allow these waves to eventually reach chromospheric heights \citep{2021_Vigeesh} and encounter their wave breaking heights, as discussed in \citet{1981_Mihalas_Toomre,1982_Mihalas_Toomre}. Analysis by \citet{2016_Hague_Erdelyi} and \citet{2009_Newington_Cally,2011_Newington_Cally} indicate that even weak vertical magnetic fields can modify the properties of AGWs irrespective of radiative damping, reflecting them back down to the lower solar atmosphere as slow mode MHD waves. These simulations seemingly demonstrate that AGWs are strongly affected by the magnetic field topology. The only observational indication of the suppression of propagating AGWs at locations of strong magnetic flux was given by \citet{2008_Straus} using high-resolution, multi-wavelength data from the Interferometric BIdimensional Spectrometer\,\citep[IBIS;][]{Cavallini_IBIS} and Michelson Doppler Imager\,\citep[MDI;][]{1995_Scherrer_Bogart_MDI}. The authors speculated this was evidence of the mode conversion of AGWs into Alfv\'{e}n waves.

To properly characterize the behavior of AGWs, high-resolution, multi-wavelength narrowband imaging of the lower solar atmosphere is necessary. The large horizontal propagation speeds associated with these modes and their potential as magnetic field diagnostics necessitates simultaneously observed multi-height velocity fluctuations derived from spectral lines that have relatively small formation height separations and are not overly sensitive to the magnetic field. Detailed studies of these waves at locations other than disk center and around more magnetic environments are necessary to provide insight into their large horizontal velocities and magnetic character. 

In the first of several papers, we revisit AGWs at quiet Sun disk center equipped with high spatial and temporal resolution, multi-wavelength ground based data in tandem with co-aligned space based data spanning the lower solar atmosphere. We employ Fourier analysis and local helioseismology techniques on derived line core Doppler velocities and intensities to probe the characteristics of AGWs. This detailed paper will illustrate the behavior of AGWs seen side by side through both intensity and Doppler velocity diagnostics for the first time. It will serve as a baseline that can be referenced for upcoming datasets exploring the behavior of AGWs at different viewing angles on the Sun. Upcoming papers will show consistent line pairs, color scales, analysis, reduction processes, and instruments to facilitate comparisons in order to better understand AGW behavior throughout the lower solar atmosphere.

While our main interest is to study propagating AGWs and compare our observations to established simulations and theory, it would be a disservice to not spend some time discussing the neighboring wave regimes. The complexity of the solar atmosphere and the relative lack of detailed AGW observations lends itself to a discussion of the full acoustic-gravity wave spectrum so that we can anchor and compare our results to previous observations. Moreover, such detailed studies are imperative to examine what information we could learn from multi-height observations with new global helioseismology projects and with the next generation solar telescopes, such as the Daniel K. Inouye Solar Telescope\,\citep[DKIST;][]{2021_Rast__DKIST, 2020_Rimmele_DKIST}. We anticipate that DKIST will provide many new multi-height observations that will help us better understand the dynamics of the solar atmosphere.

The paper is structured as follows. In Section\,\ref{sec:observations}, we discuss our 2.75 hour long co-spatial and co-temporal high-resolution, multi-wavelength ground and space based observations. We derive line core Doppler velocities and line minimum intensities for several photospheric spectral lines. In Section\,\ref{sec:waveanalysis}, we discuss the isothermal $k_{\rm h}-\nu$ diagnostic diagram and dispersion equation typically used to differentiate the various wave regimes. We use Fourier analysis to construct phase difference and coherence spectra between combinations of spectral diagnostics to investigate the behavior of AGWs, and we estimate the separation in formation height between diagnostics. In Section\,\ref{sec:discussion}, we highlight the importance of multi-height observations and use local helioseismology techniques, such as time-distance analysis, to explore the horizontal properties of AGWs. The conclusion follows in Section\,\ref{sec:conclusion}.

\section{Observations} \label{sec:observations}

\begin{figure}[htb!]
 \centering
    \includegraphics[width=\linewidth]{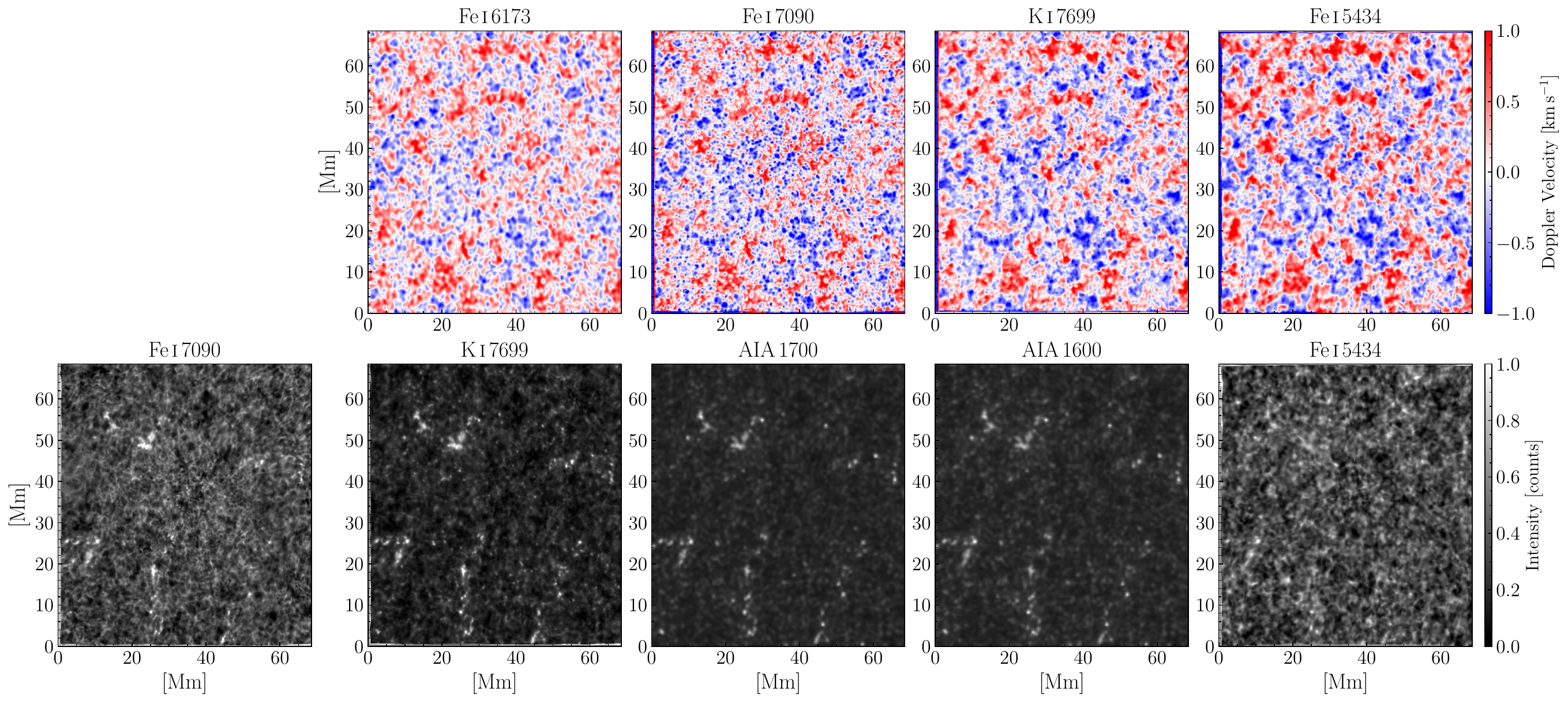}
\caption{The normalized derived line core Doppler velocity and line minimum intensity snapshots are in order of increasing average formation height. The field of view is 96\arcsec\,$\times$\,96\arcsec. Top Row: Co-temporal and co-spatial snapshots of the photospheric IBIS line core Doppler velocities and HMI Dopplergram (\ion{Fe}{1}\,6173\,\AA). Bottom Row: Co-temporal and co-spatial snapshots of the photospheric IBIS line minimum intensities and the ultraviolet AIA passbands.  \label{fig:snapshots}}
\end{figure}



\begin{deluxetable}{lccccccc}[htb!]
\tablenum{1}
\tablecaption{Observed Spectral Line Parameters \label{tab:line_information}}
\tablehead{
\colhead{} & \multicolumn{3}{c}{IBIS} & \phantom{} & \multicolumn{3}{c}{SDO\,$-$\,HMI/AIA}\\
\cline{2-4}
\cline{6-8}
\colhead{} & \colhead{\ion{Fe}{1}\,7090} & \colhead{\ion{K}{1}\,7699} & \colhead{\ion{Fe}{1}\,5434} && \colhead{\ion{Fe}{1}\,6173} & \colhead{1700} & \colhead{1600}
}
\startdata 
{Cadence [s]} & 11.88 & 11.88 & 11.88 & \phn & 12.0 & 12.0 & 12.0 \\
{${\rm g_{\rm eff} }$} & 0.0 & 1.33 & 0.0 & \phn & 2.5 & \nodata & \nodata \\
{Formation Height [km]} & 200 - 250$^{1,2}$ & 450 - 650$^{3,4,5}$ & 500 - 650$^{6,7}$ & \phn & 100 - 150$^{8,9}$ & 360$\pm 325^{10}$ & 430$\pm 185^{10}$ \\
\enddata
\tablerefs{(1) \cite{2008_Straus}; (2) \cite{2006_Janssen_Cauzzi}; (3) \cite{2017_QuinteroNoda}; (4) \cite{1986_Severino_Roberti_Marmolino_Gomez}; (5) \cite{2007_Haberreiter}; (6) \cite{2011_kneer_Bello}; (7) \cite{2010_BelloGonzalez}; (8) \cite{2011_Fleck}; (9) \cite{2014_Nagashima}; (10) \cite{2005_Fossum-Carlsson}}
\tablecomments{The formation heights listed for IBIS are average quiet Sun estimates based on the formation of the velocity signal in the line core above the base of the photosphere.}
\end{deluxetable}
We observed a quiet Sun region near disk center (110{\arcsec}, 21{\arcsec}) for 2.75 hours starting at 14:15:11\,UT on 25 April 2019 using the Interferometric BIdimensional Spectrometer \citep[IBIS;][]{Cavallini_IBIS} that was installed at the Dunn Solar Telescope (DST) in Sunspot, New Mexico along with the DST's high-order adaptive optics system \citep{2004_Rimmele_Richards_AO}. IBIS sampled the absorption line profiles of \ion{Fe}{1}\,7090\,\AA, \ion{K}{1}\,7699\,\AA, and \ion{Fe}{1}\,5434\,\AA\,with a spatial sampling of 0{\farcs}096 per pixel and an overall time cadence of 11.88\,s. The time delay between the sequential sampling of the three line profiles is 0.0, 3.26, and 7.24\,s, respectively. In addition to the narrowband data, IBIS simultaneously recorded whitelight images centered on 7200\,\AA\,\citep{2006_Cavallini_Reardon}. The Rapid Oscillations in the Solar Atmosphere instrument \citep[ROSA;][]{ROSA_Jess} also ran simultaneously, but we do not use those datasets in this paper.

To complement our ground based observations, we use space based data from the Helioseismic and Magnetic Imager\,\citep[HMI;][]{HMI_instrumental} and the lower ultraviolet Atmospheric Imaging Assembly\,\citep[AIA;][]{2012_Lemen} passbands on board the Solar Dynamics Observatory\,\citep[SDO;][]{SDO_2012}. We use HMI's Dopplergram (\ion{Fe}{1}\,6173\,\AA) to provide an independent measure of the line of sight velocities in the lower photosphere and the line of sight magnetogram for information on the lower photospheric magnetic field. The AIA\,1600\,\AA\,and AIA\,1700\,\AA\,intensity sequences are used to probe the upper photosphere and low chromosphere. Because these spectral lines are well studied, their estimated formation heights are relatively well known\,\citep[e.g.;][]{2011_Fleck, 2005_Fossum-Carlsson} and can be used to place observational constraints on our IBIS data products. The SDO data products have initially been interpolated to have a spatial sampling of 0{\farcs}6 per pixel and time cadence of 12.0\,s following Rob Rutten's SDO alignment IDL package \footnote{Rob Rutten's IDL software to co-align SDO image sequences can be \href{https://robrutten.nl/Recipes_IDL.html}{found here.}}. An overview of our observed spectral lines can be found in Table\,\ref{tab:line_information}. From here onward, we will drop the angstrom designation when discussing the wavelengths used in our study.

\subsection{Data Reduction} \label{subsec:data_reduction}
Standard flatfielding, gain, and dark calibrations were applied to both the whiteband and narrowband IBIS channels. Additional corrections for any time-dependent shifts found between the IBIS channels, prefilter transmission curves, and Fourier filtering small-scale interference fringe patterns were implemented. The spatially dependent systematic wavelength blueshift caused by the collimated mounting of the Fabry-P\'{e}rot spectrometer was also corrected. We used the nearest in time and co-spatial HMI continuum intensity images to co-align the IBIS whitelight channel and used grid images to remove optical distortions. This process corrected for any residual image motion and distortion caused by atmospheric seeing and optics without removing solar flows. We used the whitelight channel as a reference to map the narrowband, HMI, and AIA data to the same image geometry.

\subsection{Data Properties} \label{subsec:data_properties}
\begin{figure}[htb!]
    \centering
    \includegraphics[width=\textwidth]{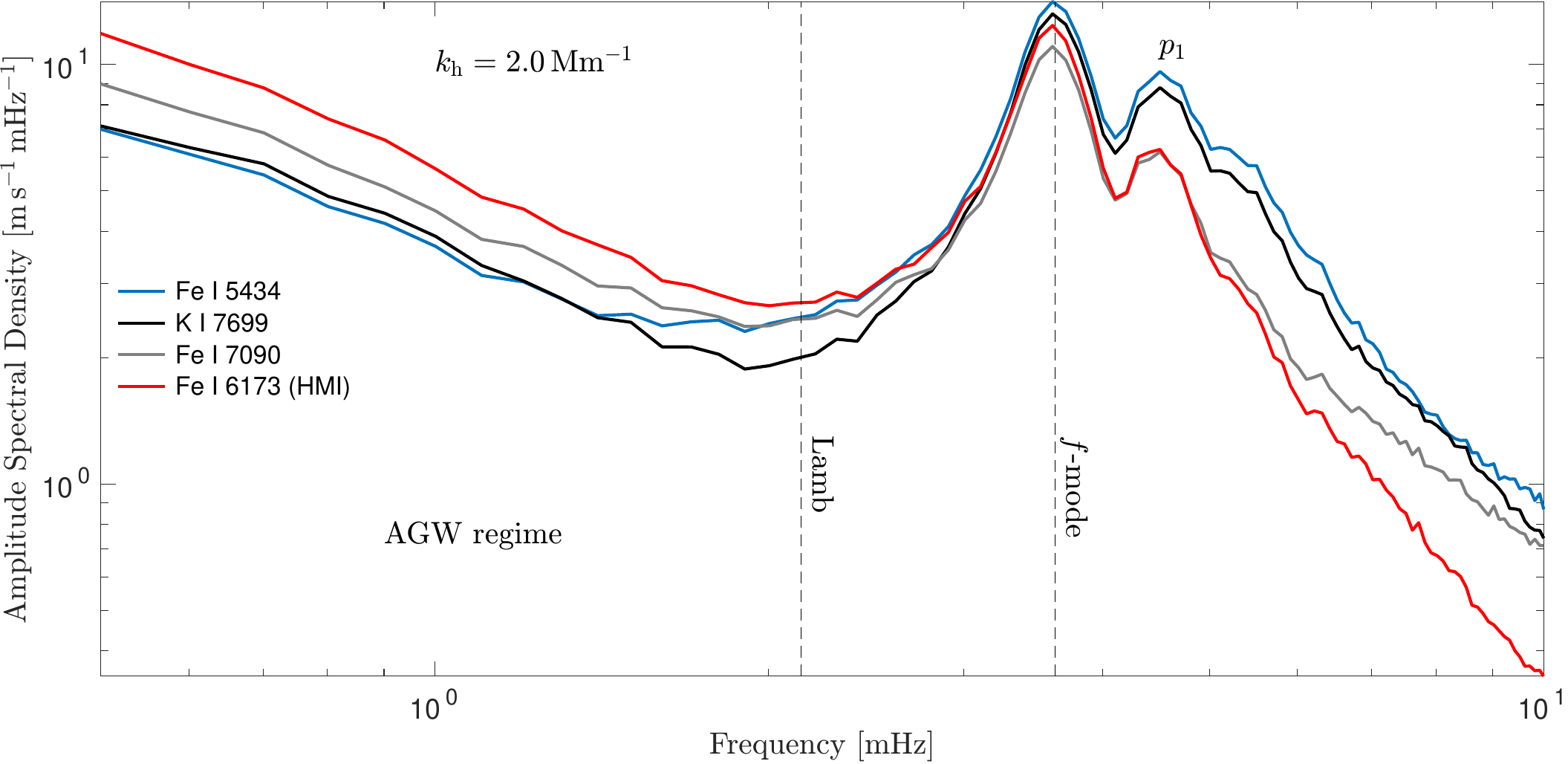}
    \caption{Velocity amplitude spectral density profiles of IBIS and HMI at a horizontal wavenumber cut of 2.0\,Mm$^{-1}$. The AGW regime at low frequencies along with the Lamb line ($\omega = ck_{\rm h}$), the $f$-mode ($\omega^2=gk_{\rm h}$), and the first $p$-mode is labelled.}
    \label{fig:amps}
\end{figure}

We compute line core Doppler velocities and line minimum intensities for each spectral line scan. We mapped the spectral data onto an evenly spaced wavelength grid and derived these data products by fitting a 2\textsuperscript{nd} order polynomial to 5 points around the line minimum. The Doppler velocities were then converted into physical units of km\,s$^{-1}$. Simultaneous snapshots of the derived line core Doppler velocities and line minimum intensities can be seen in the top and bottom row of Fig.\,\ref{fig:snapshots}, respectively.

The comparability of our observables can be demonstrated by computing velocity amplitude spectral densities as shown in Figure\,\ref{fig:amps}. We see similarities in the overall shape of the velocity amplitude profiles between our IBIS line core Doppler velocities and HMI's Dopplergram (\ion{Fe}{1}\,6173). We find a good match between the lower photospheric data given by HMI (red) and \ion{Fe}{1}\,7090 (gray) and the upper photospheric data given by \ion{K}{1}\,7699 (black) and \ion{Fe}{1}\,5434 (blue) in the evanescent regime near the $f$-mode and $p$-modes. At large frequencies (5-10\,mHz), our IBIS lines show strong velocity signals, falling off slower than the HMI Dopplergram. In this same regime, we find that the upper photospheric \ion{Fe}{1}\,5434 attains the largest amplitude while the lower photospheric HMI signal has the smallest amplitude, which is expected given their relative formation heights and the upward decrease in densities (see Table\,\ref{tab:line_information}). At low frequencies within the AGW regime, we see the opposite behavior with HMI having the highest amplitude. This might be due to the fact that \ion{Fe}{1}\,6173 is more convectively dominated as it forms lower in the photosphere.

To facilitate direct comparisons between our ground and space based observations and reduce computational time, we rebinned the IBIS data to match the spatial sampling of HMI and AIA (0\farcs6 per\,pixel). All SDO data products were ultimately interpolated to have the same cadence as IBIS. Prior to interpolation, our IBIS dataset had a Nyquist frequency of 42.1\,mHz, frequency resolution of 101\,\textmu\,Hz, Nyquist wavenumber of 45.8\,Mm$^{-1}$, and wavenumber resolution of 0.09\,Mm$^{-1}$. Afterward, the dataset had a Nyquist wavenumber of 7.3\,Mm$^{-1}$ and wavenumber resolution of 0.09\,Mm$^{-1}$. As AGWs have typical horizontal wavenumbers between 2$-$4\,Mm$^{-1}$, the rebinning process preserved the spatial scale at which we can resolve them. We also have sufficient frequency resolution to resolve these long-period oscillations. We note no quantitative differences in the results of the overall wave behavior when comparing our spatially interpolated IBIS dataset to the original as well as when we temporally interpolate our IBIS dataset to match the original cadence of HMI. Our analysis accounts for the time delay caused by the sequential sampling of the different spectral line cores. Based on the properties of Fourier transforms, the phase difference of the time lag is a linear function of frequency, so it can be added to the final azimuthally averaged phase difference spectra instead of interpolating all the time series onto one common temporal grid.

\section{Wave Analysis} \label{sec:waveanalysis}
\begin{figure}[htb!]
    \centering
    \includegraphics[width=0.7\textwidth]{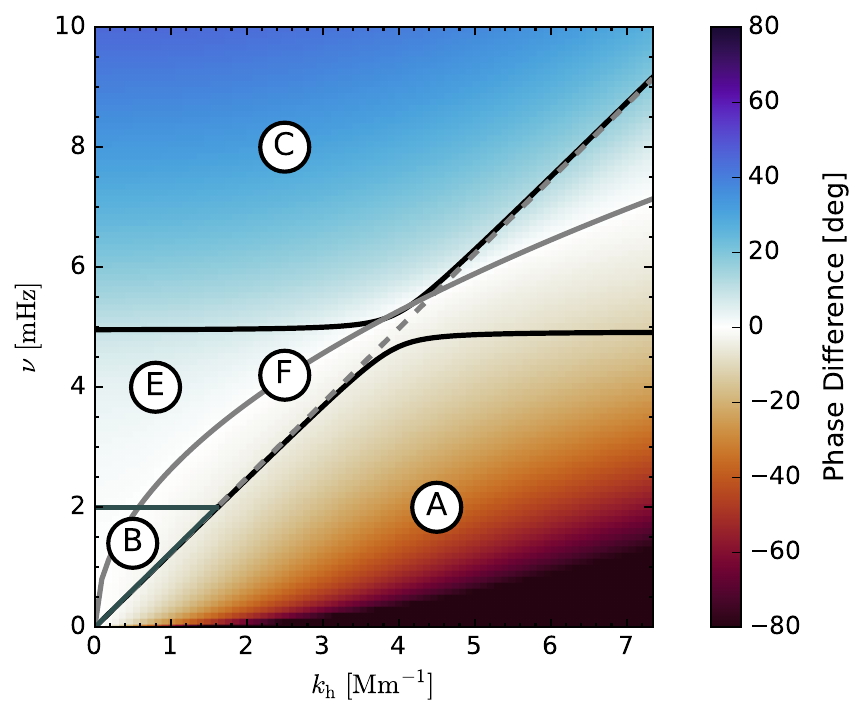}
    \caption{A theoretical phase difference spectrum calculated using Souffrin's acoustic-gravity wave theory, which describes the propagation of acoustic-gravity waves in an isothermal stratified atmosphere with constant radiative damping \citep{1966_Souffrin}, is shown on a $k_{\rm h}-\nu$ diagnostic diagram. The following values were assumed: $c_{\rm s} = 7.0$\,km\,s$^{-1}$, $\Delta z = 150$\,km, $g = 0.274$\,km\,s$^{-2}$, $\gamma = 5/3$, and $\tau_{\rm r} = 200$\,s. The following wave boundaries were determined by solving the local dispersion equation in Eqn.\,\ref{eqn:dispersion_equation}. The solid black curves denote the boundaries separating the evanescent ($k_{\rm z}^2<0$) and vertically propagating waves ($k_{\rm z}^2>0$) with AGWs located in the lower right domain. The solid, gray curved line depicts the surface gravity mode ($f$-mode). The dashed, gray line depicts the Lamb line (purely horizontal acoustic waves). Regions of interest that are frequently mentioned in this paper are labeled as follows: A for propagating AGWs; B for the wedge of evanescent waves under 2\,mHz and with horizontal wavenumbers between 0\,$-$\,1.6\,Mm$^{-1}$; C for acoustic waves; E for evanescent waves; and F for the $f$-mode.}
    \label{fig:diagnosticdiagram}
\end{figure}
We use Fourier analysis on the line core Doppler velocity (V) and line minimum intensity (I) time series to study AGWs in detail. The three-dimensional fast Fourier Transform (FFT) algorithm is used to compute phase difference and coherence spectra between combinations of spectral lines as shown in \citet{2017_Vigeesh}. The coherence ranges between zero and one, where phase differences associated with a high coherence value are considered significant. Prior to computing the FFT, we applied a running time difference on the high cadence datasets, where each image in the data cube is subtracted from the successive image, in order to remove any stationary signals. We tested various detrending and filtering methods, and the results are all quantitatively similar within the uncertainties.

The Fourier products are azimuthally averaged in the $k_{\rm x}-k_{\rm y}$ plane and illustrated on a horizontal wavenumber-frequency ($k_{\rm h}-\nu$) diagram (where $\nu = \omega / 2\pi$). The wave behavior of the acoustic-gravity wave spectrum is differentiated by solving the local dispersion equation for a compressible, gravitationally stratified isothermal medium, which is represented by
\begin{equation}
\label{eqn:dispersion_equation}
k_{\rm z}^2 = \frac{(\omega^2 - \omega_{\rm ac}^2)}{c_{\rm s}^2} - \frac{(\omega^2 - N^2)k_{\rm h}^2}{\omega^2},
\end{equation}
where $k_{\rm h}$ is the horizontal wavenumber ($k_{\rm h}^2 = k_{\rm x}^2 + k_{\rm y}^2$), $\omega$ is the angular frequency, $c_{\rm s}$ is the photospheric sound speed, $\omega_{\rm ac}$ is the acoustic cutoff frequency, and $N$ is the Brunt-V\"{a}is\"{a}l\"{a} (buoyancy) frequency. The dispersion equation separates oscillatory behavior into three distinct wave regimes: propagating acoustic waves ($k^2_{\rm z} \geq 0$) at large frequencies and large horizontal wavenumbers; evanescent or standing waves ($k^2_{\rm z} \leq 0$); and propagating AGWs ($k^2_{\rm z} \geq 0$) at small frequencies and modest horizontal wavenumbers.

We discriminate between upward propagating AGWs and acoustic waves by their contrasting phase properties. AGWs are typically associated with a negative phase difference (as a result of their orthogonal group and phase velocities). Thus, an AGW carrying energy upwards throughout the solar atmosphere will be observed with a negative phase difference\,\citep{1981_Mihalas_Toomre}. The phase difference spectrum in Fig.\,\ref{fig:diagnosticdiagram} clearly shows this defining characteristic where AGWs are displayed with negative phase differences seen in orange. As there exists a parallel relationship between the phase and group velocities of acoustic waves, a propagating acoustic wave carrying energy upward will display a positive phase difference. The following regions of interest that are mentioned throughout this paper are labeled in Fig.\,\ref{fig:diagnosticdiagram}: A for propagating AGWs; B for the wedge of evanescent waves under 2\,mHz and horizontal wavenumbers between 0\,$-$\,1.6\,Mm$^{-1}$; C for propagating acoustic waves; E for evanescent waves; and F for the $f$-mode.

\subsection{Phase Difference and Coherence Spectra} \label{subsec:phase_spectra}
We compute velocity-velocity (V\,$-$\,V), intensity-intensity (I\,$-$\,I), and intensity-velocity (I\,$-$\,V) phase difference and coherence spectra for combinations of simultaneously observed spectral diagnostics. These phase relations allow us to probe the behavior and vertical propagation of AGWs throughout the lower solar atmosphere. The first listed spectral diagnostic in the titles of the V\,$-$\,V and I\,$-$\,I phase spectra indicates the typically higher forming line. The computations were carried out in such a way as to highlight the characteristic signature of AGWs described in Section\,\ref{sec:waveanalysis}.

\subsubsection{Velocity - Velocity Phase Difference Spectra} \label{subsubsec:VV_phase_spectra}

\begin{figure}[htb!]
    \centering
    \includegraphics[width=\linewidth]{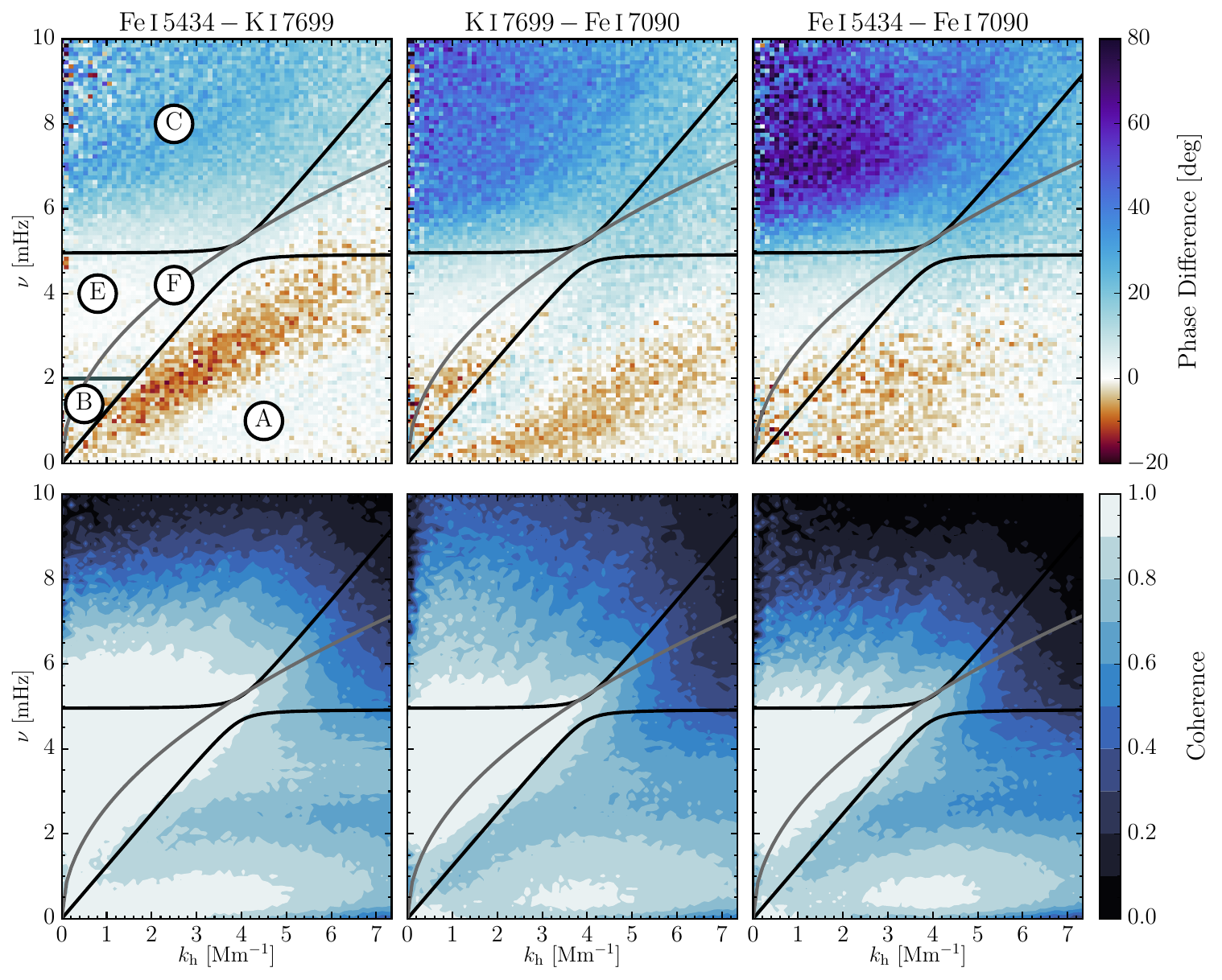}
    \caption{V\,$-$\,V phase difference (top row) and magnitude-squared coherence (bottom row) spectra derived from the photospheric IBIS line core Doppler velocities. The plots are ordered in increasing measured height difference between the diagnostics (see Table\,\ref{tab:estimated_formation_height_table}). The wave boundaries and labeled regions are the same as those shown in Figure\,\ref{fig:diagnosticdiagram}. Note the color scale which was chosen so as to not over-saturate the vertically propagating wave regimes.}
    \label{fig:azim_VVphase_IBIS}
\end{figure}

We examine the propagation of AGWs using V\,$-$\,V phase difference spectra as shown in Fig.\,\ref{fig:azim_VVphase_IBIS}, which displays the phase lag between the measured line core velocity fluctuations for combinations of IBIS spectral diagnostics.The V\,$-$\,V phase difference spectra for IBIS\,$-$\,HMI (\ion{Fe}{1}\,6173) combinations are provided in the appendix in Fig.\,\ref{fig:azim_VVphase_IBIS_HMI}.

The phase difference spectra (top row) and corresponding magnitude-squared coherence spectra (bottom row) are displayed in order of increasing formation height difference based on the measured separations calculated in Table\,\ref{tab:estimated_formation_height_table}. Following from the distinctive propagation properties in the acoustic-gravity wave spectrum, acoustic waves (Region C) propagating upward through the atmosphere will show up with a positive phase difference (blue) while propagating AGWs (Region A) carrying energy upward will show up with a negative phase difference (orange).

In Region A, we clearly detect the signature of propagating AGWs carrying energy upwards with phase differences as much as $-20\degree$. Their observable signature at low temporal frequencies ($\lesssim$\,4.5\,mHz) is prevalent for all horizontal wavenumbers with high coherence values. We see varied phase difference distributions in Region A for all IBIS combinations. In the \ion{Fe}{1}\,5434\,$-$\,\ion{K}{1}\,7699 pair, Region A shows a propagating AGW signal confined to a positive slope of some width spanning all horizontal wavenumbers for increasing frequencies with relatively high coherence values. Between 2$-$4\,Mm$^{-1}$, we see a concentration of slightly larger negative phase differences exceeding $-20\degree$. Below this signature, the phase differences are essentially $0\degree$. In contrast, the \ion{K}{1}\,7699\,$-$\,\ion{Fe}{1}\,7090 combination shows a comparable AGW signature at lower frequencies. At higher frequencies, the phase differences are positive, which is indicative of propagating AGWs carrying energy downwards. This phase difference distribution mirrors that seen in \ion{Fe}{1}\,5434\,$-$\,\ion{K}{1}\,7699.

While we detect similar overall wave behavior in Region A, we note slight differences present in the phase difference distribution between the IBIS\,$-$\,IBIS and IBIS\,$-$\,HMI combinations. IBIS\,$-$\,HMI combinations appear to have slightly larger negative phase differences. These phase difference distributions take on a defined oval shape for frequencies below 3\,mHz and horizontal wavenumbers between 1$-$6\,Mm$^{-1}$. This is in contrast to that seen in the aforementioned IBIS line combinations. However, we note that the IBIS\,$-$\,IBIS combinations that include \ion{Fe}{1}\,7090 look comparable to the IBIS\,$-$\,HMI combinations when \ion{Fe}{1}\,6173 is substituted. This behavior seems not to be related to the measured separation heights as we see this for line combinations with large and small measured formation height differences (\ion{Fe}{1}\,5434\,$-$\,\ion{Fe}{1}\,7090 versus \ion{Fe}{1}\,7090\,$-$\,\ion{Fe}{1}\,6173).

Phase analysis of our multi-height observations also provide additional insight into the acoustic-gravity wave spectrum as a whole. While we detect increasing positive phase differences with increasing formation height differences for propagating acoustic waves in Region C, this behavior does not hold for AGWs in Region A. The coherence in Region C also varies more drastically than in Region A: it decreases with increasing horizontal wavenumber and decreases rapidly with increasing frequency dropping below 0.5 at 8-9\,mHz. In general, we see phase differences close to $0\degree$ in Region E, which are expected for non-propagating evanescent waves. However, in the IBIS\,$-$\,IBIS combinations involving \ion{Fe}{1}\,7090 as well as the IBIS\,$-$\,HMI combinations with large measured formation height differences, a cluster of negative phase differences is clearly present in Region B with significantly large coherence values. These negative phase difference values are comparable with the AGW signature.

\subsubsection{Intensity - Intensity Phase Difference Spectra} \label{subsubsec:II_phase_spectra}

\begin{figure}[htb!]
    \centering
    \includegraphics[width=\textwidth]{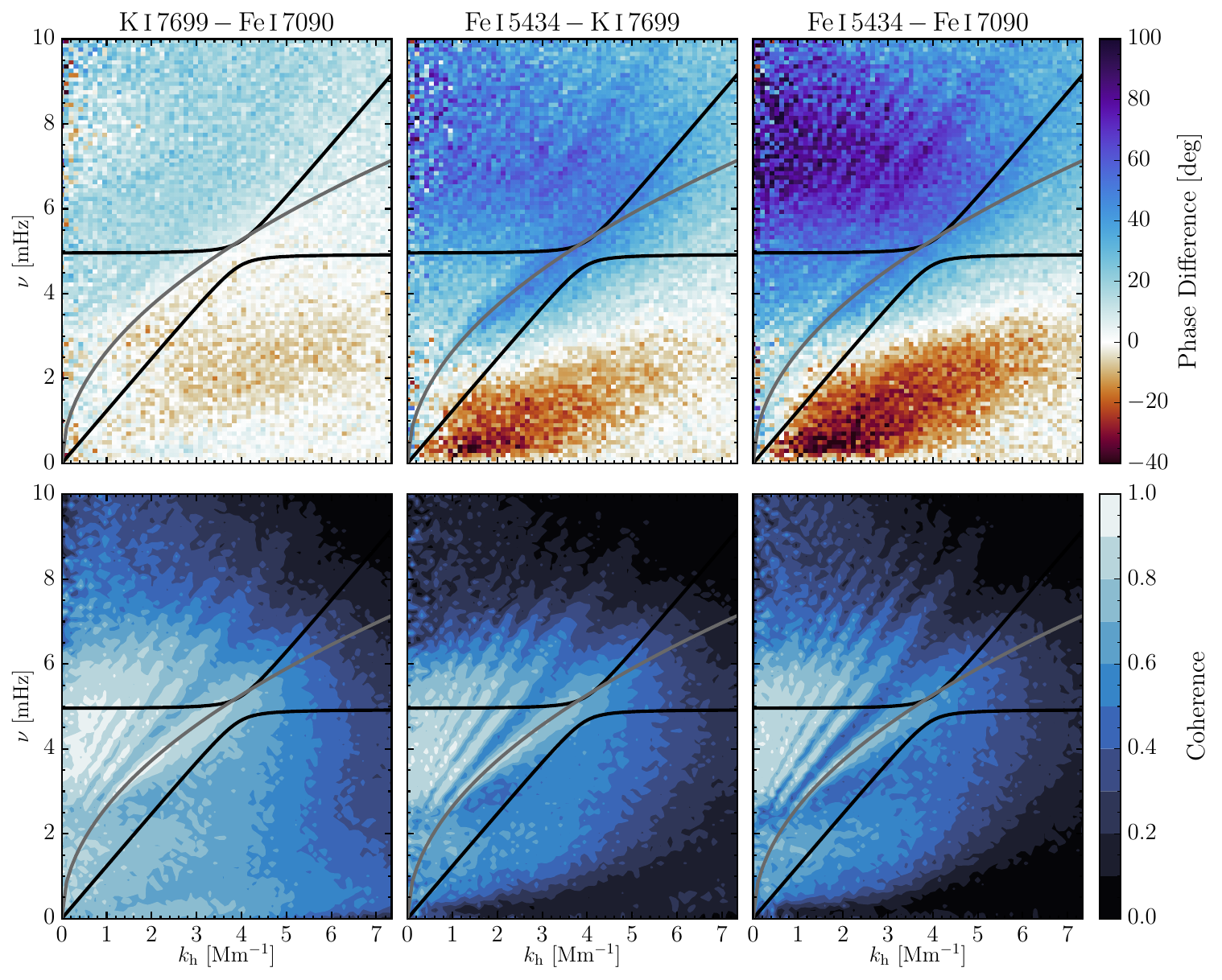}
    \caption{I\,$-$\,I phase difference (top row) and magnitude-squared coherence (bottom row) spectra derived from the photospheric IBIS line minimum intensities.  The plots are ordered in increasing measured formation height difference between the diagnostics. Note that these plots are in a different order from their V\,$-$\,V counterparts based on the estimated separation height differences indicated in Table\,\ref{tab:estimated_formation_height_table}. The color scale was chosen so as to not over-saturate the vertically propagating wave regimes.}
    \label{fig:azim_IIphase_IBIS}
\end{figure}

We examine the intensity perturbations produced by propagating AGWs using I\,$-$\,I phase difference spectra. We show phase difference spectra between IBIS\,$-$\,IBIS combinations in Fig.\,\ref{fig:azim_IIphase_IBIS}. AIA\,$-$\,AIA and AIA\,$-$\,IBIS combinations are provided in the appendix in Fig.\,\ref{fig:azim_IIphase_IBIS_AIA}. The labeled regions are consistent with the previous section, but these plots cover a greater range of phase differences. Because we are sampling the line minimum intensities that were used to derive the line core velocities, we might theoretically expect to detect similar phase difference information. However, the intensity signal is difficult to interpret as cleanly as velocity due to the fact that it is a complex agglomeration of density, temperature, and opacity effects. Radiative damping effects also need to be considered when analyzing these phase differences.

In Region A, we detect significant negative phase differences reaching $-40\degree$ congregated at horizontal wavenumbers 1$-$4\,Mm$^{-1}$ for all height combinations. In contrast to the coherence spectra shown in Fig.\,\ref{fig:azim_VVphase_IBIS}, the overall coherence is lower and decreases rapidly with increasing horizontal wavenumber and frequency. This behavior also holds for the AIA\,$-$\,IBIS combinations.

Within Region A, we notice variations and similarities present in the phase difference distribution among IBIS\,$-$\,IBIS and AIA\,$-$\,IBIS combinations. Among our IBIS diagnostics, \ion{K}{1}\,7699\,$-$\,\ion{Fe}{1}\,7090 shows overall smaller negative phase differences around $-15\degree$ mainly at higher frequencies. This is comparable to the behavior captured in the AIA\,1600\,$-$\,AIA\,1700 pair, albeit with larger coherence values. A striking phase difference distribution can be seen in the AIA\,$-$\,\ion{K}{1}\,7699 combinations, which is not found in any of the other diagnostic combinations. This negative phase difference distribution is restricted to frequencies smaller than 2\,mHz and horizontal wavenumbers 1$-$3\,Mm$^{-1}$.

Additional differences are visible when compared to Fig.\,\ref{fig:azim_VVphase_IBIS}. Within Region E, we detect mainly positive phase differences with large coherence values which might be due to radiative damping effects. Some diagnostic combinations display significant positive phase differences within Region F in addition to well-defined pseudo $p$-mode ridges within Region C. These pseudo $p$-mode ridges show increasing positive phase differences with increasing separation height and are well outlined in the coherence maps, but the coherence decreases rapidly with increasing frequency. However, these features are not present in all the line combinations, such as \ion{K}{1}\,7699\,$-$\,\ion{Fe}{1}\,7090 or AIA\,1600\,$-$\,AIA\,1700. Additionally, \ion{K}{1}\,7699\,$-$\,\ion{Fe}{1}\,7090 displays negative phase differences underneath Region F with relatively large coherence values comparable to the negative phase differences found in Region A. We note that the pseudo $p$-mode ridges are not visible within Region C for these mentioned line combinations even when truncating the color bar.

\subsubsection{Intensity - Velocity Phase Difference Spectra} \label{subsubsec:VI_phase_spectra}
\begin{figure}[htb!]
    \centering
    \includegraphics[width=\linewidth]{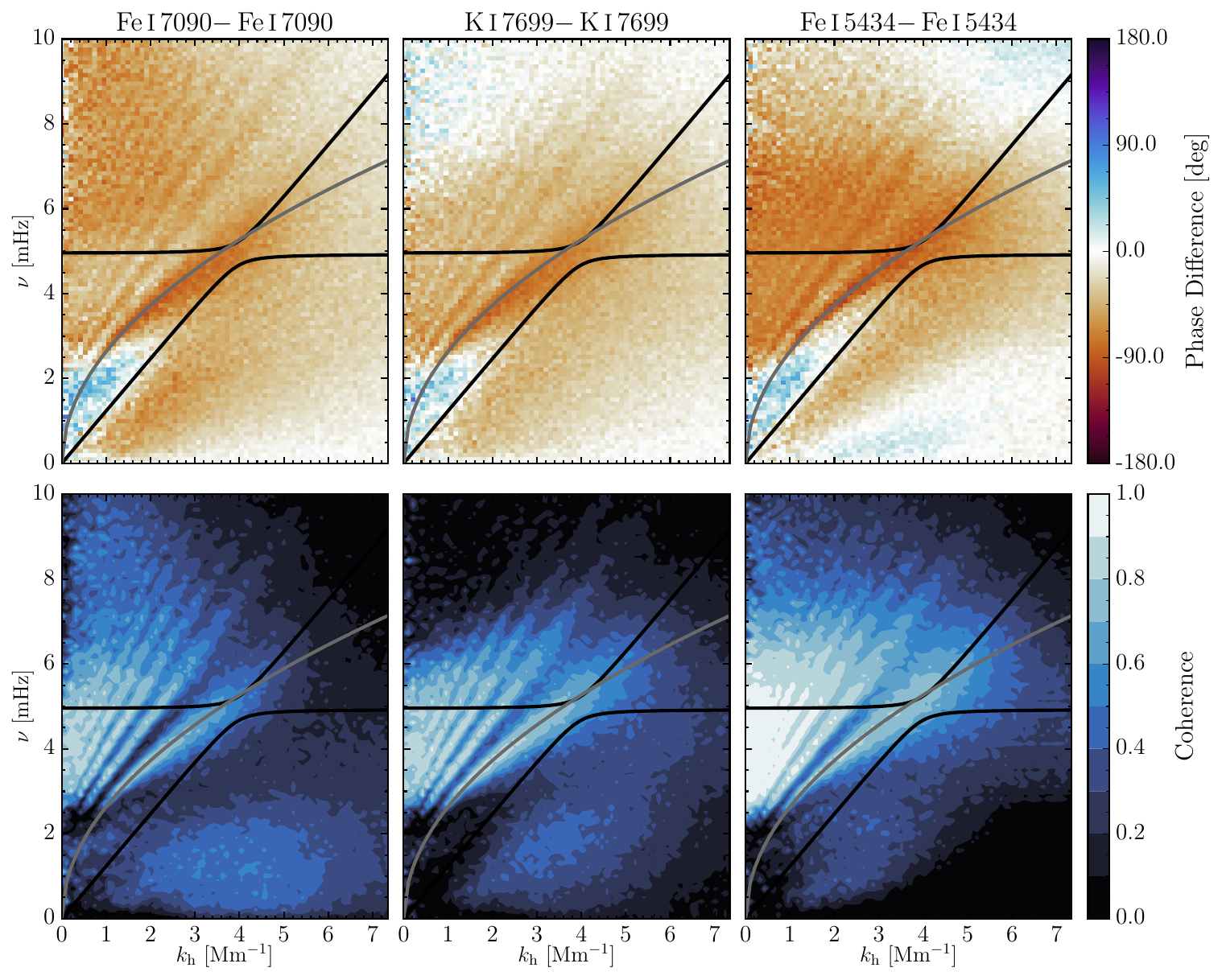}
    \caption{The phase lag between the derived photospheric IBIS line core intensity and velocity oscillations can be seen in the I\,$-$\,V phase difference (top row) and magnitude-squared coherence (bottom row) spectra. The plots are ordered in increasing average formation height as noted in Table\,\ref{tab:line_information}.}
    \label{fig:azim_VIphase_IBIS}
\end{figure}

We present the phase lag between the derived IBIS line core intensity and velocity fluctuations in the I\,$-$\,V phase difference and magnitude-squared coherence spectra in Fig.\,\ref{fig:azim_VIphase_IBIS}. For comparison, we also provide the HMI continuum and Dopplergram I\,$-$\,V phase difference and magnitude-squared coherence spectrum in the appendix in Fig.\,\ref{fig:azim_II_HMI}. I\,$-$\,V phase difference spectra hold the underlying assumption that the signals form at the same relative height in the atmosphere which might not be accurate. The phase difference spectra show distinct phase regimes with varying coherence levels corresponding to different wave behavior.

Both ground and space based diagnostics display the same overall wave behavior in the different wave regimes: negative phase differences within Region A, Region C, Region E, and Region F; and positive phase differences within Region B. However, the HMI diagnostics display overall smaller negative phase differences for Region A and C, and the pseudo $p$-mode ridges present from Region E to C are more defined. While we see little to no change in Region B between our IBIS diagnostics, this region appears to have larger phase differences in the HMI combination. The largest negative phase differences for all diagnostics appear to be associated with Region F.

In contrast to the large coherence values present in Region A in the HMI diagnostics, the overall coherence is low for all IBIS diagnostics. The largest coherence values can be attributed to Regions C, E, and F. Within Region A, the coherence levels appear to decrease with increasing average formation height. In contrast to this behavior, the coherence in Regions E and F appear to increase with average formation height. The coherence attributed to the pseudo $p$-mode ridges within Region C decreases rapidly with increasing frequency. We also see that the coherence levels present in Region B are low.

\subsection{Estimated Separation Heights Between Spectral Diagnostics} \label{subsec:formation_heights}
\begin{deluxetable}{lr@{\,--\,}lDD}[htb!]
\tablenum{2}
\tablecaption{Estimated formation height differences found between our line pairs \label{tab:estimated_formation_height_table}}
\tablehead{\colhead{ } & \multicolumn{2}{c}{Line Pair} & \multicolumn2c{$\Delta\,\phi\,(\deg)$ } & \multicolumn2c{$\Delta\,z\,(\textrm{km})$}}
\decimals
\startdata 
{\phantom{IBIS}} & \ion{Fe}{1}\,7090 & HMI & 5.9 $\pm 5.2$ & 25.6 $\pm 22.7$  \\
{\phantom{IBIS}} & \ion{Fe}{1}\,5434 & \ion{K}{1}\,7699 & 22.3 $\pm 6.3$ & 94.9 $\pm 26.4$ \\
{$\Delta\,\phi\,(\textrm{V - V})$} &  \ion{K}{1}\,7699 & \ion{Fe}{1}\,7090 & 38.7 $\pm 6.3$ & 167.9 $\pm 44.0$\\
{\phantom{IBIS}} & \ion{K}{1}\,7699 & HMI & 45.8 $\pm 7.2$ & 199.0 $\pm 50.6$\\
{\phantom{IBIS}} & \ion{Fe}{1}\,5434 & \ion{Fe}{1}\,7090 & 56.5 $\pm 9.3$ & 247.1 $\pm 68.7$ \\
{\phantom{IBIS}} & \ion{Fe}{1}\,5434 & HMI & 61.1 $\pm 9.9$ & 269.5 $\pm 80.2$ \\
\hline
{\phantom{IBIS}} & AIA\,1600 & AIA\,1700 & 14.5 $\pm 3.8$ & 65.6 $\pm 28.1$ \\
{\phantom{IBIS}} & \ion{K}{1}\,7699 & \ion{Fe}{1}\,7090 & 16.6 $\pm 6.1$ & 76.3 $\pm 40.7$\\
{\phantom{IBIS}} & \ion{Fe}{1}\,5434 & AIA\,1600 & 19.7 $\pm 6.4$ & 85.8 $\pm 32.1$\\
{\phantom{IBIS}} & \ion{Fe}{1}\,5434 & AIA\,1700 & 37.9 $\pm 7.9$ & 166.4 $\pm 52.7$\\
{$\Delta\,\phi\,(\textrm{I - I})$} & AIA\,1700 & \ion{K}{1}\,7699 & 38.2 $\pm 9.3$ & 168.7 $\pm 61.8$\\
{\phantom{IBIS}} & AIA\,1700 & \ion{Fe}{1}\,7090 & 39.7 $\pm 8.2$ & 179.6 $\pm 72.5$\\
{\phantom{IBIS}} & AIA\,1600 & \ion{Fe}{1}\,7090 & 43.7 $\pm 13.3$ & 205.1 $\pm 106.5$\\
{\phantom{IBIS}} & AIA\,1600 & \ion{K}{1}\,7699 & 45.3 $\pm 11.0$ & 206.7 $\pm 91.4$\\
{\phantom{IBIS}} & \ion{Fe}{1}\,5434 & \ion{K}{1}\,7699 & 50.6 $\pm 8.0$ & 227.5 $\pm 86.2$\\
{\phantom{IBIS}} & \ion{Fe}{1}\,5434 & \ion{Fe}{1}\,7090 & 72.1 $\pm 10.6$ & 323.6 $\pm 116.3$\\ 
\enddata
\tablecomments{The measured phase differences ($\Delta\,\phi$) are averages found in the propagating acoustic wave regime above the acoustic cutoff frequency within the range of 6$-$9\,mHz and 1$-$3\,Mm$^{-1}$. The estimated separation in formation height ($\Delta\,z$) is calculated using Eqn.\,\ref{eqn:changeinformationz}. The total phase speed for this frequency and horizontal wavenumber range is 11.6\,kms$^{-1}$.}\end{deluxetable}

To interpret the observed phase differences, we need to understand what atmospheric regions our spectral lines might sample. While we can use established quiet Sun average formation heights (see Table\,\ref{tab:line_information}), the solar atmosphere is highly corrugated, and these values are only derived from the respective line core velocity signal. The line core velocity and line minimum intensity signal might not sample the same atmospheric region, and these values would not be applicable to regions of different magnetic field strengths. For consistency, we will apply the same technique to our intensity spectral diagnostics with the aim of better understanding them. However, given the complexity of the intensity signal, these values might not accurately reflect their sampled formation height differences.

We measure the estimated separation in formation height ($\Delta\,z$) using the observed phase differences ($\Delta\,\phi$) present in the propagating acoustic wave regime (Region C) within the range 6$-$9\,mHz and 1$-$3\,Mm$^{-1}$. This range was selected to encompass the majority of the significant positive phase differences present within Region C that have relatively large coherence values. The estimated values seen in Table\,\ref{tab:estimated_formation_height_table} are calculated using
\begin{equation} \label{eqn:changeinformationz}
\Delta\,z = \frac{v_{\rm p,z} \Delta\,\phi}{2 \pi \nu},
\end{equation}
where v$_{\rm p,z}$ is the vertical phase speed and $\nu$ is the cyclic frequency. The vertical phase speed is
\begin{equation} \label{eqn:totalphasespeed}
    v_{\rm p,z} = \frac{\omega}{k_z}\approx \frac{c_{\rm s}}{\sqrt{1 - \frac{\omega_{\rm ac}^2}{\omega^2}}}.
\end{equation}
When the term involving the horizontal wavenumber is small in Eqn.\,\ref{eqn:dispersion_equation}, we get the approximate expression in Eqn.\,\ref{eqn:totalphasespeed}, which shows that the vertical phase speed can be much larger than the sound speed. Over the region in which we measure the phase differences, the vertical phase speed averages about 11.6\,km\,s$^{-1}$. We assume an acoustic cutoff frequency of 5.4\,mHz, a Brunt-V\"ais\"al\"a frequency of 4.9\,mHz, and a photospheric sound speed of 7.0\,km\,s$^{-1}$. We do not measure the separation in formation height using phase differences present within the AGW domain (Region A) as it is not immediately clear how to do so. There is greater uncertainty and more unknown variables (such as magnetic field topology and radiative damping) to take into account than there are for propagating acoustic waves.


As the HMI \ion{Fe}{1}\,6173 line and ultraviolet AIA continuum channels are well-studied spectral diagnostics, we can use them as references to constrain the estimated formation heights of our sampled IBIS fluctuations. Based on Table~\ref{tab:line_information}, the HMI \ion{Fe}{1}\,6173 represents the lowest photospheric velocity diagnostic while \ion{Fe}{1}\,5434 represents the highest photospheric velocity diagnostic. From our analysis, we find that \ion{Fe}{1}\,7090 forms in the photosphere slightly above HMI while \ion{K}{1}\,7699 forms closer to \ion{Fe}{1}\,5434.

In terms of the observed line minimum intensity fluctuations, \ion{Fe}{1}\,5434 appears to sample a higher atmospheric region than the two AIA channels by about 16$-$125\,km. In addition, the \ion{K}{1}\,7699 line minimum intensity appears to sample a lower atmospheric height than its velocity counterpart. Table\,\ref{tab:estimated_formation_height_table} shows that between the \ion{Fe}{1}\,5434\,$-$\,\ion{K}{1}\,7699 pair, $\Delta\,z\,\simeq\,$\,95\,km between their velocity signals and $\Delta\,z\,\simeq\,$\,228\,km between their intensity signals.

These $\Delta\,z$ values represent only an approximation and might not necessarily align with the differences between the established literature values. While we adopt commonly used values for the photosphere, variations in the acoustic cutoff frequency \citep{2016_Wisniewska} and the photospheric sound speed exist throughout the solar atmosphere which would alter $\Delta\,z$.

By using synthetically generated velocity spectral maps of the magnetically insensitive \ion{Fe}{1}\,5434 and \ion{Fe}{1}\,5576 lines, \citet{2020_Vigeesh_Roth} show that the AGW signature is only reliable up to a height separation of 400\,km in the photosphere and about 100$-$200\,km in the chromosphere. In other words, they find that the oblique propagation of AGWs implies that small height separations between diagnostics are necessary to obtain large coherence values. As our observables fall within this expected height difference range, we expect them to show high coherence levels. We find relatively high coherence for the V\,$-$\,V and I\,$-$\,I phase difference spectra but not for the I\,$-$\,V phase difference spectra which is not well understood. The I\,$-$\,V coherence spectra show anomalous behavior that warrants further analysis.

\section{Discussion} \label{sec:discussion}
\subsection{Noteworthy Features}
In addition to propagating AGWs, phase analysis of our multi-height observations demonstrates several noteworthy features present: (1) a varying distribution of observed negative phase differences within Region A; (2) unexpected negative phase differences present in Region B; (3) no height dependence for the phase differences in Region A; (4) distinct differences among the V\,$-$\,V and I\,$-$\,I phase difference spectra; and (5) significant phase differences in Region F and pseudo $p$-mode ridges in Region C. These observations also highlight the need to disentangle how to interpret the intensity diagnostics.

The IBIS V\,$-$\,V phase difference spectra (Fig.\,\ref{fig:azim_VVphase_IBIS}) showcase a strong variation in the distribution of negative phase differences present within Region A, which are associated with propagating AGWs carrying energy upwards. To the best of our knowledge, we have not seen such varied phase difference distributions as that seen in \ion{Fe}{1}\,5434\,$-$\,\ion{K}{1}\,7699 ($\Delta\,z\,\simeq 95$\,km) and \ion{K}{1}\,7699\,$-$\,\ion{Fe}{1}\,7090 ($\Delta\,z\,\simeq 168$\,km) previously documented. Visually, these combinations appear to be mirror images of one another. For these combinations, the propagating AGW signal is confined along a slope of some spatial scale width for all frequencies. Positive phase differences are also present in Region A, which might be an indicator of reflected AGWs carrying energy downward. Given how infrequently studied AGWs are in addition to the complexity of their modeled wave behavior, this varying distribution warrants further attention. 

The aforementioned combinations also show comparable negative phase differences to AGWs in Region B, which is typically uncharacteristic of evanescent waves. This feature seems to be evident in a couple of IBIS\,$-$\,HMI combinations as well as several I\,$-$\,I combinations, mainly involving the upper photospheric IBIS diagnostics. In contrast, Region B shows positive phase differences in all of our I\,$-$\,V phase difference spectra, the opposite phase relationship to the neighboring wave regimes, albeit with a coherence of nearly 0. This distinctive region has been previously reported in I\,$-$\,V phase difference spectra \citep[e.g.][]{1990_Deubner_Fleck_Marmolino_Severino,1992_Deubner, 1996_Deubner_Waldschik_Steffens,1999_Straus_Severino, 1998_Straus_Fleck, 2013_Severino_Straus_Oliviero_Steffen}. This feature is present in simulated V\,$-$\,V phase difference spectra with different magnetic field strength configurations by \citet{2017_Vigeesh,2019_Vigeesh} and \citet{2021_Vigeesh} and in a couple of other observed V\,$-$\,V phase difference spectra by \citet{2008_Straus} and \citet{2009_Straus_Fleck_Jeffries_Severino_Steffen_Tarbell}. The presence of this feature seems to not be related to the duration of the observed time series, not exclusive to the \ion{Fe}{1}\,7090 line core velocity signal, not related to magnetic field strength, and not associated with a specific height separation between the sampled velocity signals.

While evanescent waves present within Region E are not expected to propagate vertically ($\Delta\,\phi \simeq 0\degree$), dissipative mechanisms such as radiative damping can influence wave dynamics \citep[e.g.][]{1970_Stix,1982_Mihalas_Toomre,1991_Marmolino}. Souffrin's acoustic-gravity wave theory \citep{1966_Souffrin} notes that when radiative damping in the solar atmosphere is taken into account, the acoustic-gravity wave spectrum is altered from the rigid boundaries that are drawn in our figures \citep[see Fig.\,2 of][]{1967_Schatzman_Souffrin}. When accounting for radiative damping, waves can be broadly split into two categories: mainly progressive or mainly damped waves (which includes both AGWs and evanescent waves). For evanescent waves, the inclusion of radiative damping means that these waves are no longer purely stationary \citep{1967_Schatzman_Souffrin}. In their analysis of  I\,$-$\,V phase difference spectra, \citet{1990_Deubner_Fleck_Marmolino_Severino} suggested that these distinctive phase differences are indicative of downward propagating evanescent waves that are produced by the scattering of resonant $p$-modes, which \citet{2013_Severino_Straus_Oliviero_Steffen} confirmed after analyzing temperature gradient and opacity changes within the atmosphere.

Additionally, we find no obvious relationship between phase differences and separation height in Region A as that seen in Region C. While the positive phase differences within Region C increase with increasing height separation between the diagnostics which is expected for acoustic waves, Region A does not show a similar trend. While there seems to be a correlation between increasing separation height and phase differences present in the IBIS intensity diagnostics, this breaks down in the AIA\,$-$\,IBIS combinations. Such a trend might not be apparent due to radiative damping effects, the magnetic field strength, or even how we choose to study AGWs. Phase difference spectra mainly sample the vertical phase differences. As AGWs are known to have large horizontal velocities \citep{1981_Mihalas_Toomre}, this might not show up in our analysis. Future work is necessary to understand the observational effects of radiative damping on AGWs.

We detect differences in the distribution and magnitude of phase differences present between our I\,$-$\,I and V\,$-$\,V phase difference spectra. On average, we find larger negative phase differences as well as a more spread out distribution present in Region A in our I\,$-$\,I diagnostic combinations. Only in AIA\,$-$\,AIA and \ion{K}{1}\,7699\,$-$\,\ion{Fe}{1}\,7090 do we find smaller phase differences. Additionally, all AIA\,$-$\,\ion{K}{1}\,7699 phase difference spectra show a restricted distribution of negative phase differences below 2\,mHz in Region A, which is not evident in any other diagnostic combinations.

Propagating AGWs are believed to leave observable intensity signatures, which might theoretically show up as negative phase differences in I\,$-$\,I phase difference spectra. \citet{1981_Mihalas_Toomre} demonstrated that AGWs can have significantly larger temperature amplitudes which increase steeply with height towards their wave breaking heights in the chromosphere, where they dissipate into small-scale turbulence. It might be that these phase difference spectra are indicative of propagating AGWs that perturb atmospheric regions that either intensify or filter their intensity signal. However, there is uncertainty regarding the intensity as just a proxy for temperature due to the fact that it is heavily influenced by opacity, temperature, and density fluctuations.

The varying distribution seen in the AIA\,$-$\,\ion{K}{1}\,7699 combinations might be partly attributed to the radiative transfer properties of the spectral diagnostics. The AIA\,1600 channel is dominated by continuum emission and \ion{C}{4} while the AIA\,1700 channel only contains continuum emission\,\citep{2012_Lemen}.\cite{2017_QuinteroNoda} provides a discussion on how the atmospheric parameters encoded within the \ion{K}{1}\,7699 line core are sensitive to different layers of the atmosphere. The line minimum intensity seems to sample a lower photospheric height around where inverse granulation happens than its line core Doppler velocity which appears to be sensitive to the velocity fields in the upper photosphere. We can see similarities between the line core velocity maps of \ion{K}{1}\,7699 and \ion{Fe}{1}\,5434 while the line minimum intensity maps of \ion{K}{1}\,7699 and \ion{Fe}{1}\,7090 look similar. The measured separation heights noted in Table\,\ref{tab:estimated_formation_height_table} between \ion{K}{1}\,7699 and \ion{Fe}{1}\,7090 also affirms this as $\Delta\,z\,\simeq\,$\,168\,km between velocity fluctuations and $\Delta\,z\,\simeq\,$\,76\,km between intensity fluctuations. However, the discrepancy seen within Region A between the AIA\,$-$\,\ion{Fe}{1}\,7090 and AIA\,$-$\,\ion{K}{1}\,7699 combinations indicates some underlying radiative transfer properties or wave propagation effects that are not fully understood.

The intensity phase difference spectra also feature significantly large phase differences in Region F and pseudo $p$-mode ridges in Region C with high coherence values. We note that this behavior is not detected in the \ion{K}{1}\,7699\,$-$\,\ion{Fe}{1}\,7090 combination nor is it an effect of the truncation of the color scale. Within this particular line combination, we detect negative phase differences below Region F. The behavior present in Region F associated with the $f$-mode is not fully understood. While radiative damping might be responsible for the positive phase differences seen in Region E in intensity \citep{2001_Krijger_Rutten}, the $f$-mode is an incompressible wave that only propagates horizontally and should not be affected by radiative damping \citep{1970_Stix}. \citet{2008_Mitra-Kraev} do not observe any phase shifts for the $f$-mode in the phase difference spectra between the lower chromospheric-photospheric broadband pair G-band and \ion{Ca}{2}\,H. On the other hand, \cite{2008_Straus} find a strong signature associated with the $f$-mode's vertical energy flux. The pseudo $p$-mode ridges at high frequencies have been previously detected in intensity phase difference spectra \citep[see][]{2008_Mitra-Kraev, 2003_Rutten, 2001_Krijger_Rutten}. It is widely believed that the presence and location of this feature is the result of source resonance caused by the interference of upward and downward propagating acoustic waves and the correlated background noise which makes them more prominent in intensity \citep[e.g.][]{1999_Nigam_Kosovichev,1991_Kumar_Lu,2008_Mitra-Kraev}. Additional work needs to be done to understand not only the complexity of the acoustic-gravity wave spectrum but also the intensity signal itself. This highlights the importance of conducting more multi-height observational studies.

\subsection{Comparison with other AGW Observations, Simulations, and Theory}
We report that the heights, spatial scales, and temporal frequencies at which we detect propagating AGWs are consistent with simulations, theory, and previous observations. Direct comparisons to simulations and previous observations may be difficult given the wide range of spectral lines used (different spectral properties), the different methods used to measure velocity fluctuations (which can influence the height the fluctuations sample), and the way the AGW signature is presented visually (truncation of the color scale or color map used); nonetheless, we can discuss the overall similarities present.

We begin comparisons by focusing on the V\,$-$\,V phase difference spectra as it is the diagnostic most commonly employed to study AGWs. The phase differences present in Region A ($\Delta\,\phi\,\simeq\,-20\degree$) are in line with most previous observations. We know of two studies where larger negative phase differences have been detected: the \ion{Fe}{1}\,5434\,$-$\,\ion{Fe}{1}\,5576 phase difference spectrum in \cite{2011_kneer_Bello} (t $=$ 29.4\,min; $\Delta\,z = 190$\,km) shows values of $-40\degree$ at 2.5\,mHz, and the Mg\,b\,\textsubscript{2}\,$-$\,\ion{Ni}{1}\,6768 spectrum in \cite{2009_Straus_Fleck_Jeffries_Severino_Steffen_Tarbell} (t $=$ 12\,hr; $\Delta\,z = 600$\,km) shows values greater than $-100\degree$.

Smaller phase differences ($\Delta\,\phi\,\simeq\,-10\degree)$ with large coherence have also been identified in a 45\,s cadence HMI dataset by \cite{2014_Nagashima} (t $=$ 6.4\,hr). \citet{2014_Nagashima} generated multi-height velocity diagnostics using HMI's \ion{Fe}{1}\,6173 line to create the HMI-algorithm derived Dopplergram (${\rm z}\,\simeq\,100$\,km), line core Dopplergram (${\rm z}\,\simeq\,150$\,km), and average-wing Dopplergram (${\rm z}\,\simeq\,80$\,km). The identification of propagating AGWs in the lower photosphere using HMI allows us to confidently assume that is what we see in Region A of our IBIS\,$-$\,HMI combinations. 

Numerical simulations by \citet{2017_Vigeesh,2019_Vigeesh} and \citet{2021_Vigeesh} indicate that the magnetic field modifies the behavior of AGWs in the upper photosphere while AGWs generated in the lower photosphere should not be affected. Using the HMI line of sight magnetogram, we measure an unsigned magnetic field RMS value of 24.6\,G and median value of 4.2\,G. The magnetogram also indicates magnetically concentrated areas between $-$738\,G to 687\,G. Our disk center observations are consistent with the 0\,G and 10\,G vertical magnetic field models studied by \cite{2019_Vigeesh}. When comparing the phase difference spectra to similar heights within \cite{2019_Vigeesh}, we see slight similarities in the phase difference distributions present in Region A, in particular for the upper photospheric pair. While not shown in the paper, we checked for the effects of the lower photospheric quiet Sun magnetic field using the HMI line of sight magnetogram. We computed the phase difference spectra for pixels above and below the median magnetic field value. We did not detect a change in the overall phase difference distribution of Region A using this method for any line combinations. From this, we infer that at quiet Sun disk center, the lower photospheric magnetic field does not significantly affect the propagation of AGWs, which is in line with that reported in \cite{2019_Vigeesh}. However, we still cannot account for the wave behavior seen in some of our figures and lack upper photospheric magnetic field information. 

Intensity signatures of AGWs have not been studied in as great detail as their velocity signatures; however, previous analysis of intensity oscillations has suggested their presence within the solar atmosphere. The presence of AGWs and interference with the intensity granulation pattern at mid-photospheric heights have been previously explored \citep[e.g.][]{2004_Rutten_deWijn_Sutterlin,2006_Janssen_Cauzzi,2001_Krijger_Rutten, 1991_Komm_Mattig_Nesis,1994_Salucci_Bertello_Cavallini_Ceppatelli_Righini,2003_Puschmann}. Studies by \cite{2008_Mitra-Kraev} (t $=$ 11.87\,hr) and \cite{2004_Rutten_deWijn_Sutterlin} (t $=$ 44\,min) show the AGW signature at low frequencies with phase differences greater than $-80\degree$ between the broadband G-Band and \ion{Ca}{2}\,H observations sampled using different ground and space based instruments. This tells us that the larger negative phase differences detected in our IBIS intensity combinations are not unusual; however, our values do not reach the phase differences reported in these studies. 
When comparing our AIA\,$-$\,AIA phase difference spectrum to that seen in the 22\,s cadence ultraviolet upper photospheric and lower chromospheric 1600 and 1700 time series observed with the Transition Region and Coronal Explorer (t $=$ 3.7\,hr) shown in \cite{2003_Rutten}, which should roughly sample the same heights (and therefore features), we observe roughly similar phase differences. This indicates that even interpolating the cadence to match the faster cadence of IBIS produces similar results.

The negative phase differences present at the low temporal frequencies and horizontal wavenumbers within Region A in our I\,$-$\,V phase difference spectra have been previously attributed to AGWs \citep[see][]{1992_Deubner,1997_Straus_Bonaccini}. In the mid-photosphere around 200$-$300\,km, \cite{1997_Straus_Bonaccini} (t $=$ 4\,hr) found phase differences around $-30\degree$ corresponding to the spatial and temporal characteristics associated with AGWs. Observations by \citet{1996_Deubner_Waldschik_Steffens} looking at a 64\,s cadence time series of \ion{Ca}{2}\,8542 and \ion{K}{1}\,7699 (t $=$ 8\,hr) found phase differences in Region A near $-90\degree$ in the upper photosphere and up to $-180\degree$ in the lower chromosphere.

The theoretical framework for understanding phase relations between intensity and velocity for the acoustic-gravity wave spectrum with and without radiative damping was explored in works by \citet{1991_Marmolino}, \citet{1993_Marmolino}, and \citet{1981_Mihalas_Toomre,1982_Mihalas_Toomre}. \citet{1981_Mihalas_Toomre,1982_Mihalas_Toomre} showed that AGWs behave similarly to evanescent waves, where temperature and velocity perturbations are out of phase. For adiabatic waves, this phase difference will be $-90\degree$. Radiative damping can increase the phase differences, resulting in waves at select frequencies and horizontal wavenumbers reaching values between $-90\degree$ and $-180 \degree$.

We find that on average, our phase difference spectra display the expected theoretical out of phase relationship for AGWs. However, we detect smaller than anticipated values for the different phase regimes even taking into account non-adiabatic propagation. We also do not see the linear relationship with height in Region A reported in \citet{1996_Deubner_Waldschik_Steffens}. In fact, it appears that the phase lag between intensity and velocity for \ion{K}{1}\,7699 shows smaller phase differences for AGWs than \ion{Fe}{1}\,7090. This indicates the complexity in interpreting the intensity signature.

While our observations demonstrate qualitative agreement with previous observations, there are disagreements present when compared to the expected theoretical behavior of AGWs. Even when accounting for radiative damping using Souffrin's acoustic-gravity wave theory, we do not see similar results within Region A between our data and the expected theoretical phase difference spectrum seen in Fig.\,\ref{fig:diagnosticdiagram} besides the same sign in phase differences. The theoretical phase difference spectrum in Fig.\,\ref{fig:diagnosticdiagram} depicts a gradient with a saturated distribution of negative phase differences at small horizontal wavenumbers in Region A, which we clearly do not see in our observations even with increasing separation height between diagnostics. We also do not observationally detect such large negative phase differences in any spectral combinations. However, the theoretical modeling of the waves present in Region C lines up with what we expect to see. Our observational phase difference spectra show increasing positive phase differences with increasing separation height. Thus, these dissimilarities suggest that we need a new way to probe the behavior of AGWs in addition to the traditional $k_{\rm h}-\nu$ phase difference spectra.
This is explored in Section\,\ref{subsec:timedistance}.

\subsection{Time-distance analysis of AGWs}\label{subsec:timedistance}
The phase difference spectra seen earlier (e.g., Fig.\,\ref{fig:azim_VVphase_IBIS}) provide information regarding the vertically propagating features of AGWs, since the maps at each height are co-spatial. However, AGWs have significant horizontal motions, resulting in strong perturbations to the vertical velocity field that can be tracked in the horizontal directions. This motivates a study into these motions using various techniques commonly employed in local helioseismology\,\citep{duvall1993}.

For what follows, it is useful to review some of the theory of acoustic-gravity waves in the isothermal, non-adiabatic (and non-magnetic) case \citep{1966_Souffrin,2019_Jefferies}. A height-independent radiative damping time $\tau$ is introduced, and the dispersion relation is modified from Eqn.~\ref{eqn:dispersion_equation} to
\begin{equation}
  \label{nonad}
  k_z^2 = \pm\frac{1}{2}\left[a + \sqrt{a^2+b^2}\right],
\end{equation}
where,
\begin{eqnarray}
  \nonumber
  a&=& \frac{N^2-\omega^2}{\omega^2}k_x^2 + \frac{\omega^2-\omega_{\rm ac}^2}{c^2} - \frac{1}{1+\omega^2\tau^2}\left(\frac{N^2k_x^2}{\omega^2} - \frac{N^2\omega^2}{g^2}\right),\\ \label{nonada}
  b &=& \frac{\omega\tau}{1+\omega^2\tau^2}\left(\frac{N^2k_x^2}{\omega^2} - \frac{N^2\omega^2}{g^2}\right).
\end{eqnarray}
In these expressions, $N$ is the  Brunt-V\"ais\"al\"a frequency
\begin{equation}
  N^2 = \left(\gamma - 1\right) \frac{g^2}{c^2}.
\end{equation}
To understand the observations below, we consider a simple atmosphere with adiabatic exponent $\gamma=5/3$, sound speed $c=7\,{\rm km\,s^{-1}}$, acoustic cutoff frequency $\omega_{\rm ac}=5.3\,{\rm mHz}$, and  gravitational acceleration $g=274\,{\rm m\,s^{-2}}$. This also yields $N/2\pi\approx 5$\,mHz. The negative solution to Eqn.~\ref{nonad} describes AGWs. Eqns.~\ref{nonad}-\ref{nonada} reduce to the adiabatic solution as $\tau\rightarrow\infty$.

We are interested in an observational time-distance diagram that represents the horizontal propagation of AGWs with time. In local helioseismology, this quantity is normally computed from the inverse Fourier Transform of the velocity power spectrum. For multi-height Doppler observations, one could obtain this quantity by computing the inverse transform of the cross-spectrum, which \cite{2021_Calchetti_Jefferies_Fleck_Berrilli_etal} demonstrated. A 2-D slice through this quantity at a constant time produces a ring shape in the $x$ and $y$ direction. This ring structure captures the propagation of wave signals horizontally and provides an average of the signal between any two points a certain distance apart.

We computed the time-distance diagram in this fashion, but it is a bit noisy. Instead, we compute the time-distance diagram by temporally cross-correlating a point taken from the Doppler map at the lower height (representing the lower forming diagnostic) with the average of a concentric annulus of points from the map at the upper height. After this computation for a given annulus radius and for all spatial pixels and averaging it, we repeat for a range of radii that fit within the boundaries dictated by our field of view. For each annulus radius, about 10,000 cross-correlations are averaged. This is a higher level of averaging than what is obtained in a standard time-distance diagram and results in a stronger signal.

\begin{figure}[htb!]
  \centering
  \includegraphics[width=\textwidth]{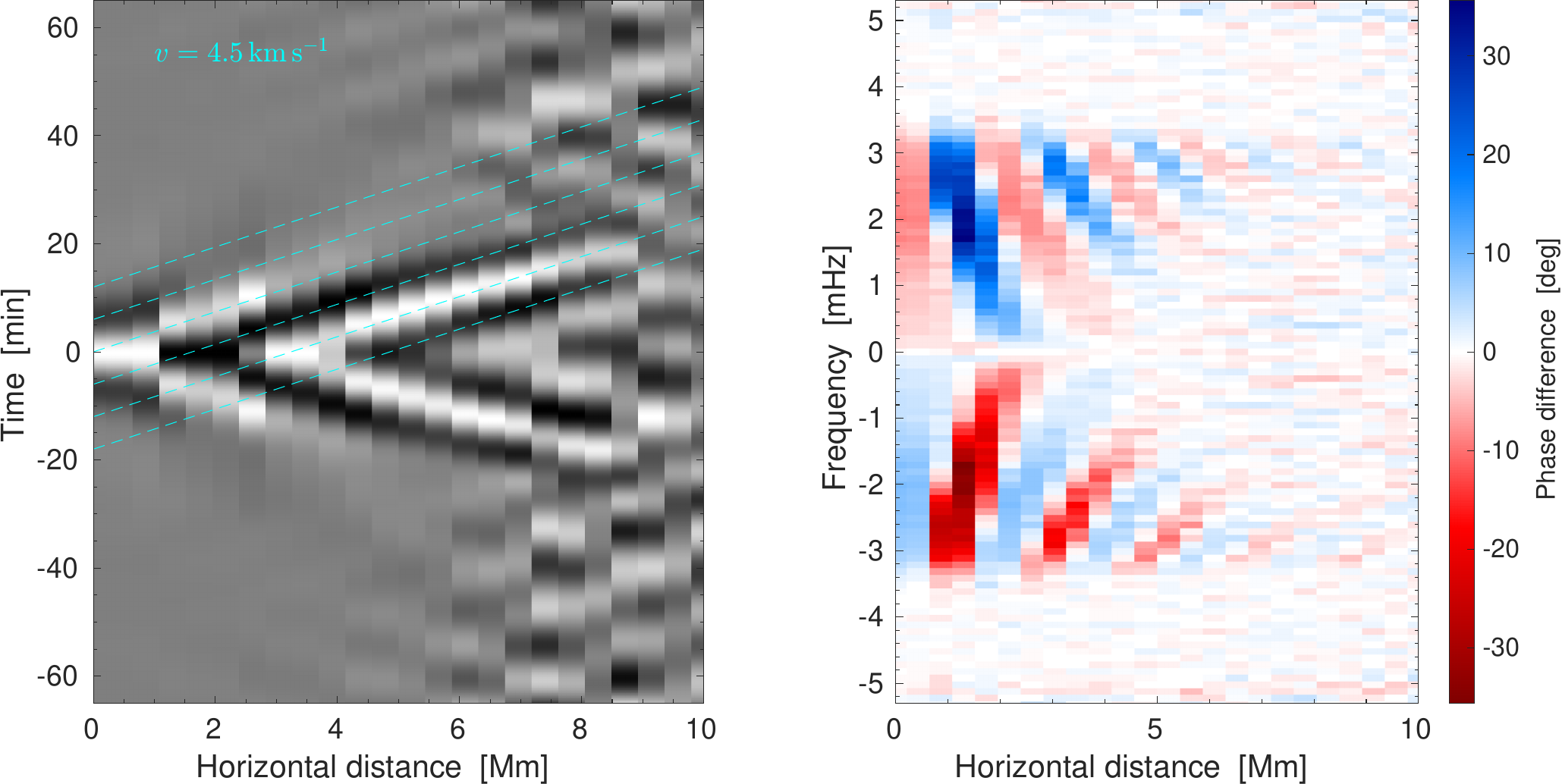}
  \caption{Strong signals from horizontally propagating AGW packets visible between the \ion{Fe}{1}\,5434 and \ion{Fe}{1}\,7090 IBIS velocity maps. (Left Panel): The temporal cross-correlation function as a function of annulus radius (distance). The gray scale is truncated to the maximum absolute value of the cross-correlation in time at each individual distance to highlight the main signal. The dashed lines indicate the nominal propagation speed. (Right Panel): The phase difference between the two heights as a function of annulus radius.}
  \label{fig:cc}
\end{figure}

We isolate the low-frequency AGWs by applying a three-dimensional Gaussian filter in both frequency and wavenumber space to the data in Fourier space prior to the computations. To avoid contamination of the acoustic waves near the Lamb line (Region E) and the $f$-mode (Region F), the filter is tapered to zero well before it reaches these regions. The resulting filter is slightly non-Gaussian with a center-of-mass at $\nu=1.17$\,mHz and $k_{\rm h}=2.28$\,Mm$^{-1}$.

We show in the left panel of Fig.\,\ref{fig:cc} a time-distance diagram computed using this method for the \ion{Fe}{1}\,7090 and \ion{Fe}{1}\,5434 line pair. We calculated V\,$-$\,V and I\,$-$\,I cross-correlations between all combinations of IBIS diagnostics. Similar results are found in all cases. The estimated height separation between these two maps is only about 250\,km (see Table\,\ref{tab:estimated_formation_height_table}); therefore, the horizontal distance is approximately the total travel distance. These results demonstrate a strong signal emanating from the low-frequency AGW packets propagating at a speed of about 4.5\,km\,s$^{-1}$, which is much slower than the local sound speed of $\sim 7$\,km\,s$^{-1}$.  After around the 8\,Mm mark, the signal becomes quite noisy, likely due to damping.

The right panel of Fig.\,\ref{fig:cc} shows the phase difference computed between the central point of the lower height (\ion{Fe}{1}\,7090) and the annulus at the upper height (\ion{Fe}{1}\,5434) for each radius. There is signal only within the frequency bandpass dictated by the filter. At the zero distance mark, which corresponds to purely vertical motion between the two layers, we replicate the phase differences ($\Delta\,\phi\,\simeq\,-15\degree$) seen in Fig.\,\ref{fig:azim_VVphase_IBIS} for this line pair combination at these approximate ($\nu, k_{\rm H}$) values, as expected. The figure shows several interesting features: (1) sign reversals of the phase differences at increasing horizontal travel distances with peak values of $\pm 35\degree$; (2) curvature of lines of constant phase difference at low frequency; and (3) a weakening of the signal at larger distances that is more rapid for the lower-frequency waves.

The observed curvature of the lines at constant phase difference can be readily explained. At any given travel distance, the higher-frequency waves within the wave packet, which have larger phase and group speeds, reach the upper height quicker and will carry their phase difference earlier than the lower-frequency waves. However, the sign reversals and the values at which they occur are more challenging to interpret.

To explore these features, we consider a very simple simulation of propagating AGWs comparable to that of \citet{2021_Calchetti_Jefferies_Fleck_Berrilli_etal}. We use a numerical 2D box set in the $x-z$ plane extending about 25\,Mm wide and a few Mm in height to propagate perturbations due to AGW packets. We prescribe the waves with positive random values for the frequency and horizontal wavenumber drawn from the distribution function of the Fourier filter that was used for our real data. We compute the (negative, downward) vertical wavenumber for each $(k_{\rm h},\nu)$ pair by solving the non-adiabatic equations. Since we extract the wavefield at two heights that are separated by 250 km, which is approximately a scale height or slightly larger than a scale height, we use an atmospheric model that does not vary in height (no stratification). We tested a stratified atmosphere model in the simulation too but found no significant difference in the results between these two layers. The values for the acoustic cutoff frequency, photospheric sound speed, gravity, buoyancy frequency, and adiabatic exponent are fixed to the aforementioned values while we vary the radiative damping timescale for different runs. The wave packets are given theoretical group velocities $\bvec{v}_{\rm g} = \partial\omega/\partial \bvec{k}$ and phase velocities $\bvec{v}_{\rm p} = \omega\bvec{k}/k^2$ computed from the atmospheric parameters \citep{1981_Mihalas_Toomre} and the wavenumber and frequency content of the considered waves. The resulting wavefield, sampled every 60\,seconds, is a linear combination of about 100 AGWs injected into the bottom left of the domain.

\begin{figure}[t!]
  \centering
  \includegraphics[width=\textwidth]{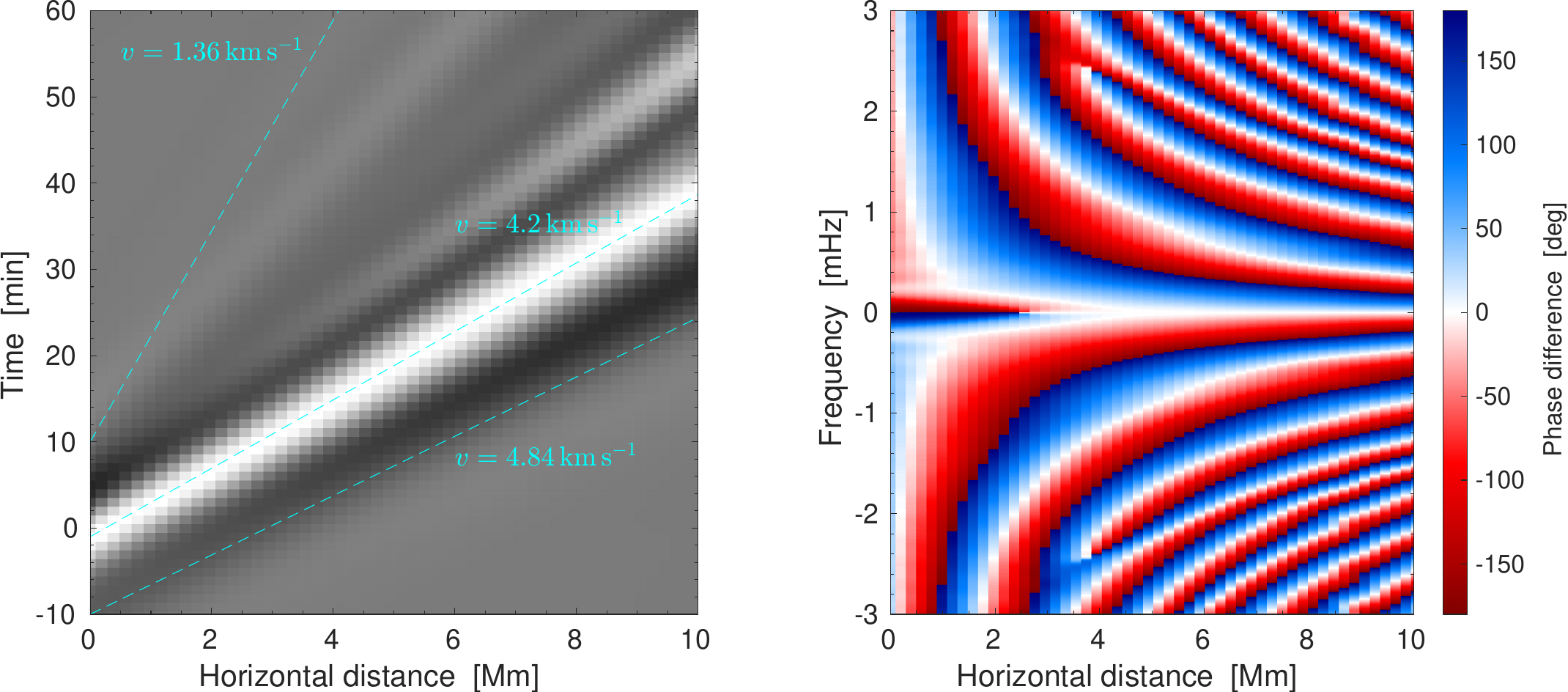}
  \caption{Time-distance and phase-difference plot for the simulated gravity waves. (Left Panel): Time-distance diagram between two layers as a function of horizontal distance. The dashed cyan lines denote the minimum/maximum range of the horizontal group velocities present in the resulting wave packet, as well as the one that best matches the main ridge. (Right Panel): Phase differences as a function of horizontal distance between the two heights separated by 150\,km. The simulated wave packet has frequency and horizontal wavenumber content in the ranges $0.4\leq\nu\leq 2.1$\,mHz and $1.2\leq k_{\rm h}\leq 3\,{\rm Mm^{-1}}$.}
  \label{fig:ccsim}
\end{figure}

To compare with our IBIS observations, we follow the same procedure and cross-correlate the wavefield at the leftmost point at the lower height with the wavefield at each successive horizontal distance at the upper height. Phase differences are also computed as a function of horizontal travel distance. Since there is no added noise in the simulation, there is no need for any additional averaging. The results are shown in Fig.\,\ref{fig:ccsim}, which can be roughly compared to the corresponding observations in Fig.\,\ref{fig:cc}. One can observe faint evidence of all the individual waves that make up the wave packet in the simulation, which show up as straight lines at different slopes. The range of group velocities is indicated in the figure by the dashed cyan lines. However, this feature is not seen in the IBIS results, which may imply some filtering mechanism present in the Sun, since the largest group speeds are for the waves with the highest frequencies. 

Indeed, we discover that the maximum of the cross-correlation in the observational time-distance diagram clearly corresponds to the horizontal group speed of the wave packets, as expected. The simulated AGWs have group speeds ranging from about 2\,km\,s$^{-1}$ to about 5\,km\,s$^{-1}$, with only a few of those with the largest frequencies attaining the highest speeds that are seen in the IBIS data. The group speed increases with frequency and is only very weakly dependent on the atmospheric values used, suggesting that the signal in Fig.~\ref{fig:cc} is dominated by the AGWs in the filter with the higher-frequency content. The low-frequency waves seem to be damped out of the cross-correlation. Finally, because we only inject waves at one location and do not consider downward propagating waves, we do not see the negative time branch signal in Fig.\,\ref{fig:ccsim} that is seen in Fig.\,\ref{fig:cc}.

The simulated phase differences have some similarities with the IBIS observations. At zero horizontal separation, we find phase differences of comparable magnitude to the data in the frequencies of interest. This agreement is only present when the radiative damping time used in the model is set to around 60\,seconds. In the adiabatic case, which corresponds to very large damping times, we find much larger phase differences that do not agree with the IBIS observations. These results make sense, since for the waves in the filter, $\omega\tau\leq 1$, and damping plays a significant role. This indicates the importance of studying and including damping effects when analyzing the propagation of AGWs. Radiative damping will reduce the vertical wavenumber and the vertical group velocity of the wave packets, and the phase differences are then propagated between heights more slowly.

The sign changes at increasing distances are also present within the model. However, we find that the phase differences wrap at values of $\pi$ instead of the smaller extrema seen in Fig.\,\ref{fig:cc}. Intuitively, the simulation results make sense as the phase differences should increase to their extrema as the waves propagate throughout the atmosphere. The cutoff of around $\pm 35^\degree$ in our data is perplexing and indicates some missing physics in the model. The damping time is fixed for both heights instead of increasing with height as is expected \citep[see][]{1982_Mihalas_Toomre}; however, that would not affect the results seen here. We also cannot replicate the rapid diminishing of the phase differences with distance present at low frequencies.

\section{Conclusions} \label{sec:conclusion}
Phase analysis of our multi-height space and ground based data from IBIS and SDO provided a window into the complex dynamics present in this observed quiet Sun disk center region, where we find propagating AGWs carrying energy upwards throughout the lower solar atmosphere. Using Fourier spectral analysis to construct phase difference and magnitude-squared coherence spectra in addition to local helioseimology techniques, we investigated both the vertical and horizontal properties of AGWs. The $2.75$\,hour long high-resolution, multi-wavelength IBIS time series provides sufficient frequency resolution to resolve these long-period oscillations (roughly 4 to 16\,min) while the addition of SDO data provided supplemental height information.

To understand the future diagnostic potential of AGWs and what information could be retrieved from a phase difference spectrum, we need to observationally understand their wave behavior in both Doppler velocity and intensity diagnostics. We find (1) propagating AGWs for all spectral diagnostic pairs and different height separations comparable with previous observations, theories, and simulations; (2) the distribution and magnitude of phase differences for AGWs varies depending on the sampled diagnostic; (3) the horizontal propagation properties of AGWs can be inferred from time-distance and phase difference diagrams; and (4) multi-height observations highlight the complexity of the solar atmosphere. 

For all height combinations, we detect significant negative phase differences in the low temporal frequency domain associated with propagating AGWs carrying energy upward. As previously detected, the signature of propagating AGWs is visible up to the lower chromosphere sampled by AIA. On average, the observed negative phase differences at these spatial and temporal scales are consistent with theory, simulations, and prior observations. However, we find that that the behavior seen observationally disagrees with that in Souffrin's acoustic-gravity wave theory for AGWs. Even though observationally these waves appear to have enough momentum to overcome radiative damping in the lower solar atmosphere, our work shows that we need to better understand the role radiative damping plays in the behavior of AGWs and how we might understand it through intensity diagnostics.

Our study also highlights the varied phase difference distributions and magnitudes present within the AGW domain for various V\,$-$\,V and I\,$-$\,I diagnostic combinations that have not been observed previously. In general, we find larger negative phase differences present between the I\,$-$\,I combinations albeit with smaller coherence than their velocity counterparts. Additionally, both the intensity and velocity diagnostics show similar overall wave behavior for AGWs: negative phase differences indicating upwards propagation. Distinctive phase difference distributions are visible in the V\,$-$\,V combinations of \ion{Fe}{1}\,5434\,$-$\,\ion{K}{1}\,7699 and \ion{K}{1}\,7699\,$-$\,\ion{Fe}{1}\,7090 and the I\,$-$\,I combinations between AIA\,$-$\,\ion{K}{1}\,7699. We also note significant positive phase differences at temporal frequencies and horizontal wavenumbers typically associated with AGWs in the V\,$-$\,V combination between \ion{K}{1}\,7699\,$-$\,\ion{Fe}{1}\,7090, which might imply the reflection of propagating AGWs carrying energy upward back down to the lower photosphere.

These line combinations also sample atmospheric regions where the behavior of AGWs are believed to be influenced by the magnetic field. While not shown in the paper, a quick analysis performed by masking pixels based on the median strength of the lower photospheric quiet Sun magnetic field defined using HMI's line of sight magnetogram showed that even for strong magnetic fields, we detect no significant changes to the overall propagation of AGWs in the velocity maps. These results are consistent with the wave behavior found in \citet{2019_Vigeesh}'s weak field magnetic runs of 0\,G and 10\,G. We infer that at least near quiet Sun disk center, the lower photospheric magnetic field does not significantly affect the generation and propagation of AGWs.

Additionally, our results enable us to comment on the sampled heights of the line core Doppler velocity and line minimum intensity fluctuations used in this study. Using a select frequency and horizontal wavenumber range in the propagating acoustic wave regime in Region C, we measured the separation in formation heights between diagnostics. We find that the \ion{Fe}{1}\,5434 line minimum intensity signal seems to sample higher atmospheric regions than the AIA passbands, and the \ion{K}{1}\,7699 line minimum intensity and line core velocity signals probe visibly different atmospheric heights.

We remind the reader that our phase difference spectra mainly provide insight into the vertical perturbations induced by propagating AGWs. As AGWs tend to have a significantly larger horizontal component (thousands of km versus one to two hundred km of separation in the vertical direction for our spectral diagnostics), we are missing out on a detailed characterization of these modes by only studying their vertical motions. In order to analyze the strong horizontal signatures expected of AGWs, we compute time-distance and phase difference diagrams as a function of horizontal displacement, which in our case is the total travel distance for these waves. The time-distance diagram for the \ion{Fe}{1}\,5434\,$-$\,\ion{Fe}{1}\,7090 velocity combination shows AGWs propagating with an approximate horizontal group speed of 4.5\,km\,s$^{-1}$, which is in line with values reported in \citet{1981_Mihalas_Toomre}. The corresponding phase difference plot shows frequency-dependent phase differences at various horizontal displacements. We can replicate the approximate vertical phase difference for this diagnostic pair for zero horizontal distance which corresponds to purely vertical motions. However, our simple simulation of about 100 AGWs only partially explains this observed behavior, and reasonable agreement is only seen if the radiative damping time is 60\,s. Our simulated time-distance diagram shows ranges of horizontal group velocities not seen in our observations. This indicates that the signal in our data is mainly dominated by high-frequency AGWs and that the low-frequency AGWs are damped out of the picture.

The complexity of the acoustic-gravity wave spectrum is clearly seen through our observations. In addition to the behavior of the AGWs, we find negative phase differences comparable to AGWs in Region B (strongly indicative of reflected evanescent waves), an out of phase relationship for Region B present in all of our I\,$-$\,V phase difference spectra, and significant phase differences present in Region F associated with the $f$-mode and prominent pseudo $p$-mode ridges in Region C in our intensity diagnostics.

As demonstrated by this work, multi-line observations provide a wealth of diagnostic potential regarding oscillations and features present in the solar atmosphere, which can even have impacts on local helioseimology measurements\,\citep[see][]{2022_Zhao}. This highlights the importance of including the measurement of multiple spectral lines spanning the atmosphere in upcoming synoptic networks, such as the next-generation GONG network\,\citep[ngGONG;][]{2019_Hill_ngGONG}. We also look forward to future DKIST observations that will explore the dynamics of the solar atmosphere and potentially allow us to better understand the observed AGW behavior. Because IBIS was dismantled at the DST in 2019, there is a need for similar narrowband imaging spectroscopic instruments able to make resolved measurements of photospheric lines. In particular, once available, the Visible Tunable Filter \citep[VTF;][]{2012_Kentischer_DKIST, 2020_Rimmele_DKIST}, may fill in this gap with its diffraction-limited imaging spectroscopy and spectropolarimetry. 



In this paper, we revisited the quiet Sun disk center to study the propagation of AGWs using both Doppler velocity and intensity diagnostics validating past results with current results. We hope that the inclusion of the intensity diagnostics alongside the velocity diagnostics motivates new 3D MHD simulations to better understand how to interpret the I\,$-$\,I and I\,$-$\,V phase difference spectra and the effects of radiative damping. A detailed observational characterization of the properties and propagation of AGWs needs a thorough investigation into their behavior in the presence of magnetic fields and horizontal propagation characteristics. Our main goal is to study AGWs at various viewing angles on the Sun for a greater understanding of their properties and behavior. In future papers, we plan to fill in the gap present in the knowledge of AGWs by studying their behavior when viewed obliquely near the solar limb in order to investigate their horizontal properties and in the vicinity of active regions to explore the effect of the magnetic field strength and orientation. We will use the analysis presented in this paper as a reference to better understand how these different environments influence the behavior of AGWs.

\begin{acknowledgments}
Data in this publication were obtained with the Dunn Solar Telescope facility, which is operated by New Mexico State University with funding support from the National Science Foundation and the state of New Mexico. We would like to thank the DST staff for all of their help and taking these observations, in particular Doug Gilliam. The HMI and AIA data used in this publication are courtesy of NASA's SDO and the HMI and AIA science teams. O.V. and J.J. acknowledge support from NASA under grant 80NSSC18K0672. O.V. is also supported under NSF grant 1936336.
\end{acknowledgments}

\facilities{Dunn(IBIS), SDO(HMI,AIA)}

\software{CMasher \citep{2020_JOSS_cmasher}}

\appendix

\section{Additional Phase Difference Spectra} \label{appendix:additional_phases} 
\begin{figure}[!ht]
    \centering
    \includegraphics[width=\linewidth]{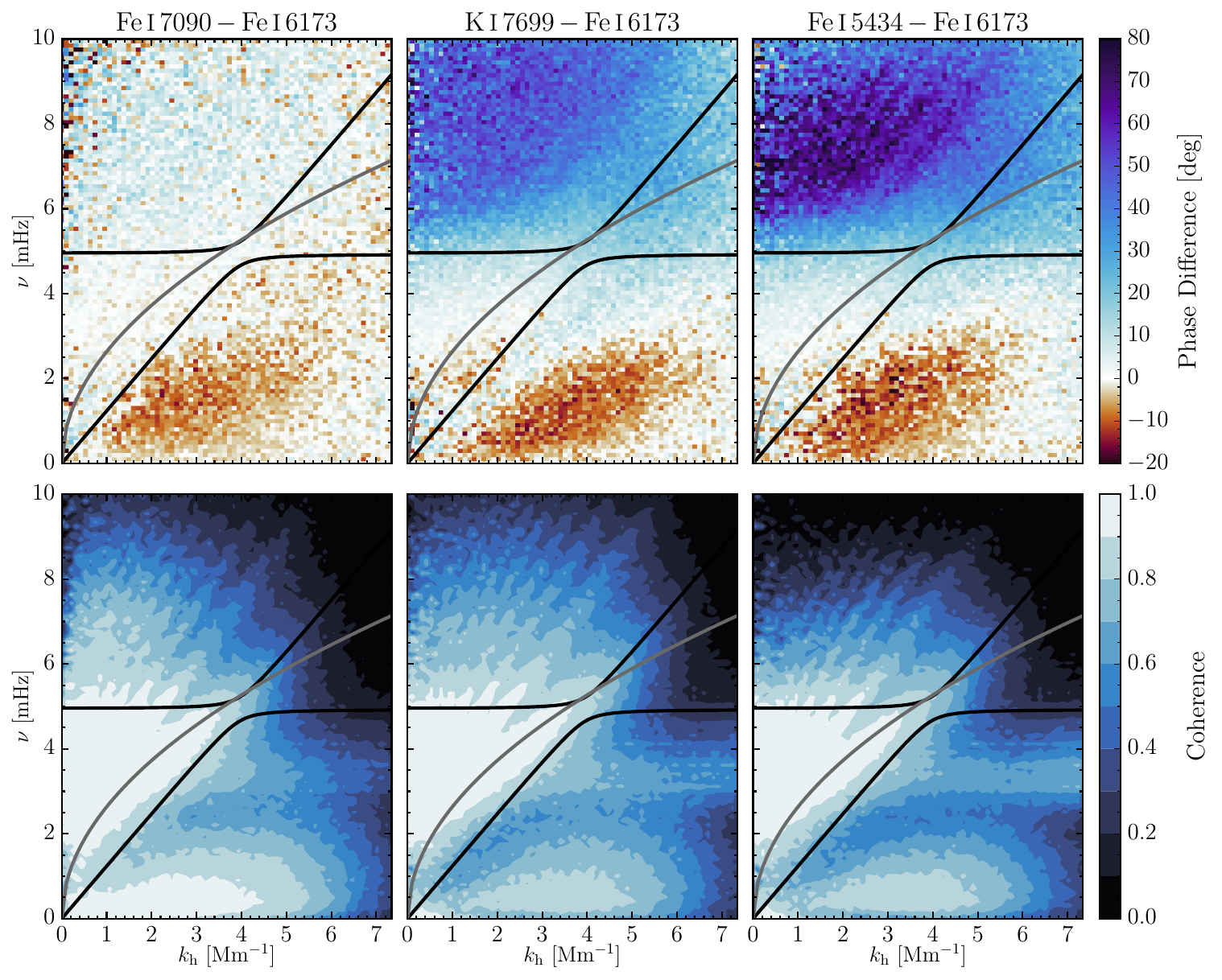}
    \caption{V\,$-$\,V phase difference (top row) and magnitude-squared coherence (bottom row) spectra showing combinations between the photospheric IBIS line core Doppler velocities and the HMI Dopplergram (\ion{Fe}{1}\,6173). The plots are in order based on the measured height separations calculated in Table\,\ref{tab:estimated_formation_height_table}.}
    \label{fig:azim_VVphase_IBIS_HMI}
\end{figure}

\begin{figure}[!ht]
    \centering
    \includegraphics[width=\linewidth]{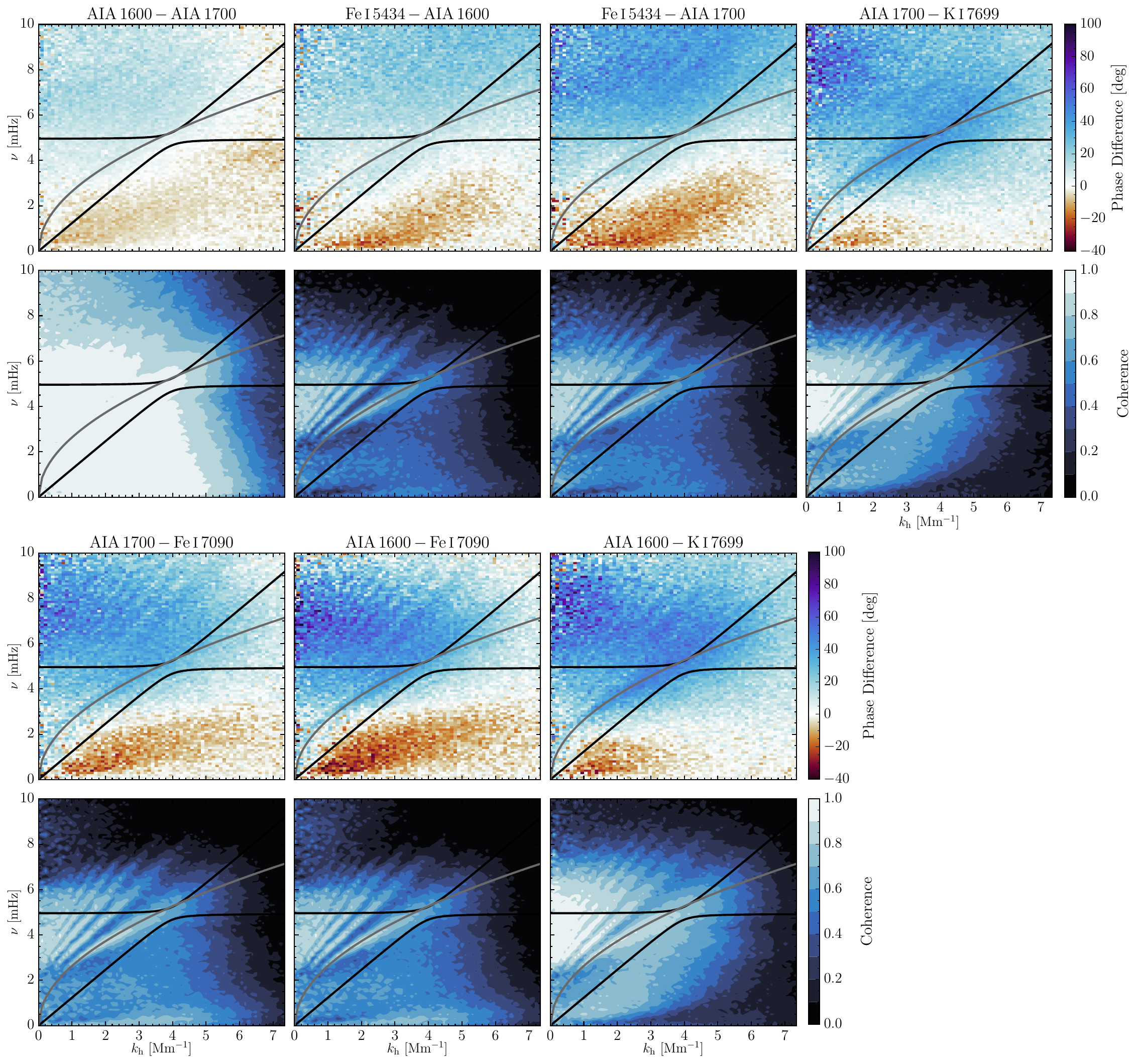}
    \caption{I\,$-$\,I phase difference (top row) and magnitude-squared coherence (bottom row) spectra  showing combinations between the photospheric IBIS line minimum intensities and the AIA passbands. The plots are in order based on the measured height separations in Table\,\ref{tab:estimated_formation_height_table}.}
    \label{fig:azim_IIphase_IBIS_AIA}
\end{figure}

\begin{figure}[!ht]
    \centering
    \includegraphics[width=\linewidth]{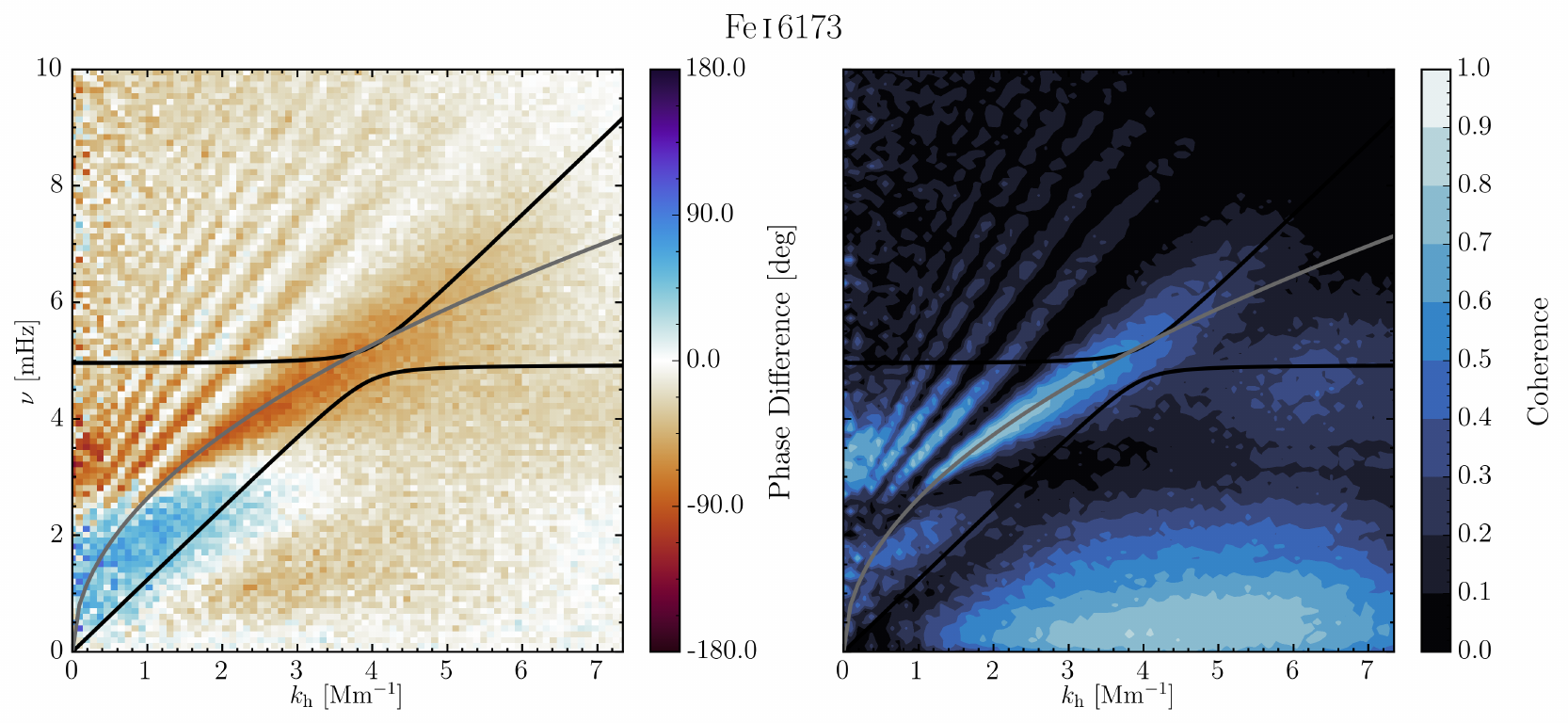}
    \caption{I\,$-$\,V phase difference and magnitude-squared coherence spectrum for the HMI (\ion{Fe}{1}\,6173) continuum and Dopplergram data.}
    \label{fig:azim_II_HMI}
\end{figure}

\clearpage
\bibliographystyle{aasjournal}
\bibliography{AGWbib}



\end{document}